\documentclass[aps,prb,groupedaddress,reprint,longbibliography]{revtex4-2}

\pdfoutput=1

\usepackage{graphicx}
\usepackage{amsmath,amsfonts,amssymb,mathtools}
\usepackage{color}
\usepackage[dvipsnames]{xcolor}
\usepackage{subfigure}
\usepackage{physics}
\usepackage{dsfont}
\usepackage{hyperref}
\usepackage{booktabs}
\usepackage{url}
\usepackage{leftidx}
\usepackage{textcomp}
\usepackage{gensymb}
\usepackage{threeparttable}
\usepackage{rotating}
\usepackage{esvect}
\usepackage{booktabs}
\usepackage{marvosym}
\usepackage{lmodern}
\usepackage[version=4]{mhchem}

\definecolor{emerald}{rgb}{0.31, 0.78, 0.47}
\definecolor{blue(ncs)}{rgb}{0.0, 0.53, 0.74}

\hypersetup{
     colorlinks=true,
     linkcolor=blue(ncs),
     filecolor=blue,
     citecolor=emerald,      
     urlcolor =blue(ncs),
}
	    
\usepackage[official]{eurosym}
\definecolor{lightgray}{rgb}{0.9,0.9,0.9}	    
\definecolor{green}{rgb}{0,0.5,0}
\definecolor{red}{rgb}{1,0,0}
\definecolor{blue}{rgb}{0,0,0.5}

\long\def\symbolfootnote[#1]#2{\begingroup%
\def\thefootnote{\fnsymbol{footnote}}\footnotetext[#1]{#2}\footnotemark[#1]\endgroup}

\usepackage{chngcntr}

\DeclareMathAlphabet{\pazocal}{OMS}{zplm}{m}{n}

\newcommand{\bo}[1]{\boldsymbol{#1}}
\newcommand{\id}{\mathds{1}}
\newcommand{\D}{\mathrm{d}}
\newcommand{\f}[2]{\frac{#1}{#2}}

\begin{document}

\title{Thermodynamic transitions and topology of spin-triplet superconductivity: Application to UTe$_2$}

\author{Henrik S.~R{\o}ising}
\author{Max Geier}
\author{Andreas Kreisel}
\author{Brian M.~Andersen}

\affiliation{Niels Bohr Institute, University of Copenhagen, DK-2200 Copenhagen, Denmark}

\date{\today}

\begin{abstract}
The discovery of unconventional superconductivity in the heavy-fermion material UTe$_2$ has reinvigorated research of spin-triplet superconductivity. We perform a theoretical study of coupled two-component spin-triplet superconducting order parameters and their thermodynamic transitions into the superconducting state. With focus on the behavior of the temperature dependence of the specific heat capacity, we find that two-component time-reversal symmetry breaking superconducting order may feature vanishing or even negative secondary specific heat anomalies. The origin of this unusual specific heat behavior is tied to the non-unitarity of the composite order parameter. Additionally, we supply an analysis of the topological surface states associated with the different possible spin-triplet orders: single-component orders host Dirac Majorana surface states in addition to possible bulk nodes. A second component breaking time-reversal symmetry gaps these surface states producing chiral Majorana hinge modes. DFT+$U$ band-structure calculations support that these topological phases are realized in ${\rm UTe}_2$ when introducing weak superconducting pairing. Our topological analysis suggests measurable signatures for surface-probe experiments to acquire further evidence of the superconducting pairing symmetry.
\end{abstract}

\maketitle

%
\section{Introduction}
\label{sec:Intro}
%
Obtaining a versatile platform with topologically protected surface states and/or persistent superconducting surface currents is of high current priority within the condensed matter physics community. These desires have naturally focused attention on spin-triplet (odd-parity) superconductivity where both properties are relevant due to nontrivial winding, and potentially also time-reversal symmetry breaking (TRSB) of the superconducting order parameter. The exploration of materials exhibiting these features and their unusual response to electromagnetic fields pose exciting research directions with relevance for robust quantum computing with emergent non-abelian anyons in the form of Majorana quasiparticles~\cite{NayakEA08}.

The material UTe$_2$ is a new candidate for topological spin-triplet superconductivity~\cite{RanScience}. In this heavy-fermion compound, superconductivity sets in at $T_c = 1.5$-$2$~K depending on the method of sample preparation~\cite{Thomas2021,Rosa2022}.  Emergence of unconventional spin-triplet pairing in UTe$_2$ is supported by several experimental facts including exceedingly large upper critical magnetic fields well beyond the Pauli-limit for spin-singlet order~\cite{RanScience,Aoki2019,Ran2019}, a modest Knight shift upon entering the superconducting state~\cite{RanScience,Nakamine2021,Tokunaga2019,Matsumura2023}, and re-entrant field-induced superconductivity~\cite{RanScience,Knebel2019}. In addition, a non-zero polar Kerr effect at $T<T_c$ signals spontaneous TRSB by the superconductivity~\cite{HayesEA21}, even though this property has recently been challenged by follow-up experiments~\cite{Ajeesh2023}. The existence of TRSB is consistent with the existence of an anomalous normal component of the conductivity found by surface microwave impedance measurements~\cite{Bae2021}, and TRSB may also be a supporting ingredient for the generation of chiral Majorana modes at step edges in UTe$_2$~\cite{MadhavenEA20}.

At present the pairing symmetry of UTe$_2$ is not agreed upon, and it remains open which irreducible representation (irrep) of the $D_{2\mathrm{h}}$ point group the superconducting condensate prefers. Restricting the discussion to odd-parity irreps in the presence of strong spin-orbit coupling (SOC) singles out the A$_{\mathrm{u}}$, B$_{1\mathrm{u}}$, B$_{2\mathrm{u}}$, and B$_{3\mathrm{u}}$ irreps as possible candidates for the order parameter symmetry~\cite{SigristUeda91,HillierEA09}. Accidental degeneracies further allows for TRSB combinations thereof, for example $\mathrm{B}_{3\mathrm{u}}+i\mathrm{A}_{\mathrm{u}}$, which additionally constitutes a rare example of a non-unitary superconducting order.

In principle, the gap symmetry in the material can be determined by detailed measurements of the momentum-dependence of the superconducting gap as done in other unconventional superconductors~\cite{Kreiselrev}, however, these measurements are very challenging given the small energy scales. At present, basically all allowed candidates and their pairwise complex combinations are being considered, but the recent discussion has largely been focused on $\mathrm{B}_{3\mathrm{u}}+i\mathrm{A}_{\mathrm{u}}$ or one of the B$_\mathrm{u}$ irreps. In this regard, a recent experimental study of the temperature- and field-orientation-dependence of the magnetic penetration depth concluded that only the two-component TRSB $\mathrm{B}_{3\mathrm{u}}+i\mathrm{A}_{\mathrm{u}}$ phase appears consistent with the data due to its nodal structure~\cite{KotaEA23}. This is in contrast to several other experiments advocating for one of the single-component B$_{\mathrm{u}}$ gap symmetries, including, e.g., recent ultrasound~\cite{Theuss2023}, field-dependent specific-heat measurements~\cite{SLee2023}, and scanning tunneling microscopy (STM) experiments~\cite{Seamus2023}. A single component condensate is consistent with specific heat measurements on high-quality samples reporting a low-temperature power law tail and a single transition that does not split under uniaxial strain~\cite{Rosa2022,Cairns2020,Girod2022,AokiEA22}.

To a large extent, the discussion of single- versus multi-component TRSB superconductivity and the associated conflicting experimental evidence parallels that of recent developments in the understanding of the superconducting phase of Sr$_2$RuO$_4$. For instance, muon spin relaxation measurements on this material indicate TRSB in the superconducting state, with a transition temperature that splits off from $T_c$ under uniaxial strain~\cite{Luke98,GrinekoEA20}. However, thermodynamic probes~\cite{DeguchiEA04,Li2021,LiEA22,GrgurEA23} and SQUID microscopy~\cite{MuellerEA23} do not observe any sign of a second transition, thus casting doubts on the two-component scenario. In fact, many probes are straightforwardly explained by a single superconducting $d_{x^2-y^2}$-wave order parameter~\cite{Mackenzie2017,Kivelson2020,RomerEA19,RoisingEA19,ASR2,Andersen2023}.

Here, motivated by the conundrum between the conflicting evidence for single- versus two-component superconductivity in UTe$_2$, we investigate the thermodynamic transition into a TRSB non-unitary spin-triplet superconductor. Specifically, we focus on the specific heat capacity and its property upon entering non-unitary TRSB phases. We find that the behavior of the temperature dependence of the specific heat capacity in the case of a two-component TRSB superconducting order may feature vanishing or even negative secondary transition anomalies. We show how the origin of such unusual specific heat behavior is tied to the non-unitarity of the composite order parameter. This result may help reconcile conflicting evidence for single- versus two-component superconductivity in UTe$_2$. 

We further describe the topology and corresponding anomalous boundary excitations of the superconducting phases relevant for UTe$_2$. These results provide additional signatures distinguishing the orders in experiments probing the surface excitations. We find that a pure superconducting order with $\mathrm{B_{3u}}$ symmetry is a second-order topological nodal superconducting phase \cite{SimonPRB2022} hosting Majorana Dirac cones \cite{QiPhysRevLett2009May, ChiuEA16} on the surface and a Majorana flat band at hinges. Breaking time-reversal symmetry with a weak admixture with $\mathrm{A_u}$ symmetric superconducting order generically gaps the Majorana Dirac surface cones and turns the flat hinge band into a chiral Majorana mode \cite{ReadGreen00}. For an approximately equal mixture of $\mathrm{B_{3u}}$ and $\mathrm{A_u}$ symmetric orders the system may transition into a Weyl superconducting phase with Fermi arcs of Bogoliubov quasiparticles \cite{MengPhysRevB2012Aug}. A pure order with $\mathrm{A_u}$ symmetry is a fully gapped, strong topological superconductor with Majorana Dirac surface states \cite{QiPhysRevLett2009May}. Here, a small admixture of $\mathrm{B_{3u}}$ symmetric superconducting order turns the system into a second-order topological superconducting phase with chiral Majorana modes on hinges. In our analysis, we emphasize the consequences for the $(0, -1, 1)$ surface that is experimentally relevant for ${\rm UTe}_2$~\cite{MadhavenEA20}. 

To support that these topological phases are realized in ${\rm UTe}_2$ when including superconducting pairing of the respective symmetry, we calculate the bandstructure using DFT+$U$ {\it ab initio} calculations. For a relevant range of moderate Hubbard repulsion $U$ where the bandstructure is metallic as observed experimentally, we find that the Fermi surface has a sheet that can be deformed into a sphere and a  cylinder without crossing any time-reversal invariant momenta. Since the cylinder by symmetry always encloses an even number of time-reversal invariant momenta, we consider only the spherical pocket as relevant for the strong topological phases discussed here. This motivates our model of a spherical Fermi surface used in the topological analysis. Calculating a symmetry-based indicator \cite{GeierPhysRevB2020Jun} from the DFT+$U$ bandstructure, we find that the previously discussed topological phases are indeed realized when introducing weak superconducting pairing of the respective symmetry in the ${\rm UTe}_2$ bandstructure.
 
The paper is organized as follows. In Sec.~\ref{sec:FreeEnergy} we present a general Landau free-energy analysis of two coupled one-dimensional (1D) spin-triplet order parameters. This section introduces the different allowed mutual structures of the two triplet irreps. Sec.~\ref{sec:SpecificHeat} contains a general discussion of the specific heat capacity and the thermodynamic anomalies at the critical temperatures of the two active components. Next, in Sec.~\ref{sec:UTe2} we turn to the particular case of UTe$_2$ and discuss its electronic structure, the thermodynamic superconducting transitions, and its topological properties including the surface states arising both from the bandstructure and the different possible superconducting order parameters. Finally, Sec.~\ref{sec:Conclusion} provides a general discussion and our conclusions. 

%
\section{Free energy of coupled spin-triplet order parameters}
\label{sec:FreeEnergy}
%
A single-band spin-triplet superconductor is characterized by the vector order parameter $\vec{d}(T,\bo{k})$ in the convenient Balian--Werthamer basis, $\bo{\Delta} = (\Vec{d}\cdot \vec{\sigma})i\sigma_2$, such that $\Vec{d}$ transforms as a vector under combined spin and spatial rotations~\cite{BalianWerthamer63}:
\begin{equation}
\bo{\Delta} = 
 \begin{bmatrix}
        \Delta_{\uparrow \uparrow} & \Delta_{\uparrow \downarrow} \\
        \Delta_{\downarrow \uparrow} & \Delta_{\downarrow \downarrow}
    \end{bmatrix} = 
 \begin{bmatrix}
        -d_x+id_y & d_z \\
        d_z & d_x + id_y
    \end{bmatrix}.
\label{eq:OrderParameter}
\end{equation}
The superconducting gaps are given by~\cite{Sigrist05, Ramires_2022}
\begin{equation}
\lvert \Delta_{\sigma} \rvert^2 = \lvert \vec{d} \rvert^2 + \sigma \lvert \vec{d}^{\ast} \times \vec{d} \rvert,
\label{eq:SpinSplitGaps}
\end{equation}
and are spin-split for so-called \emph{non-unitary states} which are characterized by $\lvert \vec{d}^{\ast} \times \vec{d} \rvert \neq 0$. 

Consider two competing triplet orders, respectively associated with symmetry-distinct 1D irreps of the corresponding point group and hence generally on-setting at two distinct critical temperatures, $T_{c1}$ and $T_{c2}$. The phase diagram can be mapped out using Ginzburg--Landau theory. As such, we retain all symmetry-allowed terms to quartic order involving two complex vector order parameters $\vec{d}_1$ (onset at $T_{c1}$) and $\vec{d}_2$ (onset at $T_{c2}$), resulting in the free energy density
\begin{widetext}
\begin{equation}
\begin{aligned}
\pazocal{F}[\vec{d}_1,~\vec{d}_2] &= \alpha(T) \lvert \vec{d}_1 \rvert^2 + \beta_1 \lvert \vec{d}_1 \rvert^4 + \beta_2 \lvert \vec{d}_1^{\ast} \cross \vec{d}_1 \rvert^2 + \tilde{\alpha}(T) \lvert \vec{d}_2 \rvert^2 +  \tilde{\beta}_1 \lvert \vec{d}_2 \rvert^4 + \tilde{\beta}_2 \lvert \vec{d}_2^{\ast} \cross \vec{d}_2 \rvert^2  \\
&\hspace{20pt} + \gamma_1 \left[ (\vec{d}_1 \cdot \vec{d}_2^{\ast})^2  + (\vec{d}_1^{\ast} \cdot \vec{d}_2)^2 \right] + \gamma_2 \left[ (\vec{d}_1 \cdot \vec{d}_1)(\vec{d}_2^{\ast} \cdot \vec{d}_2^{\ast}) + (\vec{d}_1^{\ast} \cdot \vec{d}_1^{\ast})(\vec{d}_2 \cdot \vec{d}_2) \right] \\
&\hspace{20pt} + \gamma_3 \lvert \vec{d}_1 \rvert^2 \lvert \vec{d}_2 \rvert^2 + \gamma_4 (\vec{d}_1 \cdot \vec{d}_2^{\ast}) (\vec{d}_1^{\ast} \cdot \vec{d}_2) + \gamma_5 (\vec{d}_1^{\ast} \cdot \vec{d}_2^{\ast}) (\vec{d}_1 \cdot \vec{d}_2),
\end{aligned}
\label{eq:FreeEnergy1}
\end{equation}
\end{widetext}
where $\vec{a}\cdot \vec{b} = \vec{a}^{T}\vec{b}$. This theory has nine quartic coefficients, reflecting the enhanced complexity in having multiple possible scalar contractions of three-dimensional (3D) vectors, i.e., via both the scalar product and the anti-symmetric cross product, in contrast to the case of scalar (singlet) orders.

For the theory of Eq.~\eqref{eq:FreeEnergy1} to be bounded from below we require $\beta_1,~\tilde{\beta}_1 > 0$. Further coefficient magnitude and sign criteria will be required to guarantee thermodynamic stability, as exemplified in a specific instance below. As usual, the quadratic coefficients are assumed to go negative below the respective critical temperatures, $\alpha(T < T_{c1}) < 0$, and $\tilde{\alpha}(T < T_{c2}) < 0$.

Considering for reference Eq.~\eqref{eq:FreeEnergy1} in the case of $\gamma_j = 0$ for $j \in \lbrace 1, \dots, 5 \rbrace$, the two components decouple. If $\beta_2 < 0$ (resp.~$\tilde{\beta}_2 < 0$) the component $\Vec{d}_1$ (resp.~$\Vec{d}_2$) itself becomes non-unitary. However, from a microscopic evaluation of the quartic coefficients in the absence of magnetic fields, $\beta_2$ (resp.~$\tilde{\beta}_2$) is positive semi-definite and determined by Fermi surface average of the form factors of $\Vec{d}_1$ (resp.~$\Vec{d}_2$) to the fourth power~\cite{ScheurerEA20, WagnerEA21}. Still, it can be argued that residual magnetic interactions can stabilize a single-component non-unitary order parameter~\cite{FernandesEA13, WeiEA22}. For an example of how microscopic evaluations of the Ginzburg--Landau coefficients restrict the a priori possible phases of the phenomenological theory, we refer to Appendix~\ref{app:Intermezzo}.

\subsection{Minimization with simple ans{\"a}tze}
\label{sec:Minimization}
We consider next the theory of Eq.~\eqref{eq:FreeEnergy1} in the special case in which the two order parameter components are parameterized by their amplitudes ($D_1$, $D_2$), two real unit vectors ($\hat{d}_1$, $\hat{d}_2$) and a relative, complex phase ($\varphi \in [0, \pi/2]$), i.e., 
\begin{equation}
\vec{d}_1 = D_1 \hat{d}_1,~\mathrm{and}~\vec{d}_2 = D_2 e^{i\varphi} \hat{d}_2.
\label{eq:Ansatze}
\end{equation}
In other words, we assume the constituents $\Vec{d}_1$ and $\Vec{d}_2$ to be unitary, but non-unitarity can still be induced in a coexistence phase in which $\lvert \hat{d}_1 \times \hat{d}_2 \rvert \neq 0$, and $\varphi \in (0, \pi/2]$. We emphasize that we are here concerned with possible non-unitarity in a coexsistence phase of two symmetry-distinct orders. Note that a single-component order parameter, such as $\vec{d} = \Delta_0 (k_z, -ik_z, 0)^{\mathrm{T}}$, is sufficient for non-unitarity~\cite{SigristUeda91, Ramires_2022}. This order, however, displays trivial heat capacity features in the context of the analysis in Sec.~\ref{sec:SpecificHeat}, see also Appendix~\ref{app:SingleNonUnitary}. Here, we pursue the two-component non-unitary scenario motivated by its possible relevance for UTe$_2$.

When the above ans{\"a}tze are inserted into Eq.~\eqref{eq:FreeEnergy1} the free energy reduces to a form familiar from a scalar theory analogue~\cite{Lee2009}, which is straightforwardly minimized analytically:
\begin{equation}
\begin{aligned}
    \pazocal{F}[D_1,D_2,\hat{d}_1\cdot \hat{d}_2,\varphi] &= \alpha(T) D_1^2 + \tilde{\alpha}(T) D_2^2 + \beta_1 D_1^4 \\ &\hspace{10pt} + \tilde{\beta}_1 D_2^4 + \kappa D_1^2 D_2^2, \\
\end{aligned}
\label{eq:FreeEnergy2}
\end{equation}
with 
\begin{equation}
    \kappa \equiv 2\cos{(2\varphi)}  \big[ (\hat{d}_1 \cdot \hat{d}_2)^2 \gamma_1 + \gamma_2 \big] + \gamma_3 + (\hat{d}_1 \cdot \hat{d}_2)^2 (\gamma_4+\gamma_5).
    \label{eq:Kappa}
\end{equation}
In addition to the positive definiteness imposed on $\beta_1$ and $\tilde{\beta}_1$, we must also impose $4\beta_1 \tilde{\beta}_1 > \kappa^2$ (seen by requiring positive eigenvalues of the quartic form matrix associated with the free energy potential) to ensure thermodynamic stability.

In Eq.~\eqref{eq:FreeEnergy2}, the dependence on $(\hat{d}_1 \cdot \hat{d}_2)^2$ and $\varphi$ only enters through the cross term $\kappa$, and minimization of this term gives the four possible coexistence phases, as controlled by the three parameters $\gamma_1$, $\gamma_2$, and $\nu \equiv \frac12(\gamma_4 + \gamma_5)$: $\kappa_{\mathrm{A}} = \gamma_3 - 2\gamma_2$, $\kappa_{\mathrm{B}} = \gamma_3 + 2\gamma_2 + 2\gamma_1$, $\kappa_{\mathrm{C}} = \gamma_3 + 2\gamma_1 + 2\gamma_2 + 2\nu$, and $\kappa_{\mathrm{D}} = \gamma_3 - 2\gamma_1 - 2\gamma_2 + 2\nu$, 
\begin{figure}[bh!t]
	\centering
	\includegraphics[width=\linewidth]{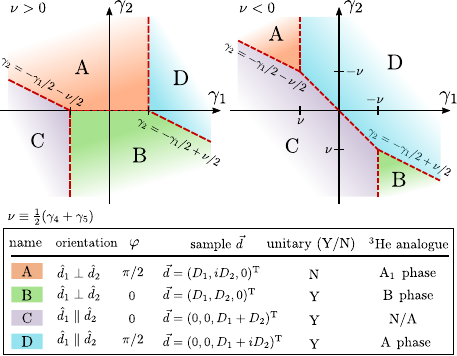}
    \caption{Theoretical phase diagrams controlled by the parameters $\gamma_1$, $\gamma_2$, and $\nu \equiv \frac12(\gamma_4 + \gamma_5)$, of Eq.~\eqref{eq:FreeEnergy1} with the ans{\"a}tze $\vec{d}_1 = D_1 \hat{d}_1$ and $\vec{d}_2 = D_2 e^{i\varphi} \hat{d}_2$. The four coexistence phases, labelled A, B, C, and D, are stabilized in the regimes of corresponding label color in the plane spanned by $\gamma_1$ and $\gamma_2$, with $\nu > 0$ in the left panel and $\nu < 0$ in the right panel. Sample $\vec{d}$ vectors and analogue phase realized in superfluid $^3$He are listed in the text box below the phase diagrams.}
	\label{fig:PhaseDiagram}
\end{figure}
with the phases for both $\nu > 0$ and $\nu < 0$ summarized in Fig.~\ref{fig:PhaseDiagram}. Given quartic coefficients satisfying $4\beta_1 \tilde{\beta}_1 > \kappa^2$ in any of the four phases, $D_1$ and $D_2$ are found by minimzing the remaining theory, given the $\kappa$'s above, over the amplitudes, resulting in 
\begin{equation}
\begin{aligned}
D_1^2 &= \max\Big\lbrace \frac{\kappa \tilde{\alpha} - 2\tilde{\beta}_1 \alpha}{4\beta_1 \tilde{\beta}_1 - \kappa^2},0 \Big\rbrace, \\
D_2^2 &= \max\Big\lbrace \frac{\kappa \alpha - 2\beta_1 \tilde{\alpha}}{4\beta_1 \tilde{\beta}_1 - \kappa^2},0 \Big\rbrace.
\end{aligned}
\end{equation}
In the four cases above, coexistence of the components require $D_1, D_2 > 0$, i.e., $\kappa_j \tilde{\alpha} > 2\tilde{\beta}_1 \alpha$ and $\kappa_j\alpha > 2\beta_1 \tilde{\alpha}$ for $j \in \lbrace \mathrm{A}, \mathrm{B}, \mathrm{C}, \mathrm{D} \rbrace$. As seen explicitly by the entrance of $\alpha$ and $\tilde{\alpha}$, these latter two requirements are in general temperature dependent. 

As summarized in Fig.~\ref{fig:PhaseDiagram}, three of the coexistence phases have (experimentally realized) analogues in superfluid $^3$He. The phase we label A, characterized by $\hat{d}_1 \perp \hat{d}_2$ and $\varphi = \pi/2$, is the only non-unitary phase. A sample order parameter for this phase is $\Vec{d} = (D_1,iD_2,0)^{\mathrm{T}}$, which explicitly breaks the symmetry between $\ket{\uparrow \uparrow}$ and $\ket{\downarrow \downarrow}$ since $\lvert \Delta_{\uparrow \uparrow} \rvert = \lvert -d_x + id_y\rvert \neq \lvert d_x + id_y \rvert = \lvert \Delta_{\downarrow \downarrow} \rvert$. This resembles the A$_1$-phase of~$^3$He, which is stabilized by an external magnetic field~\cite{AmbegaokarEA73} and can be verified through a crisp heat capacity double transition~\cite{Halperin76}. In contrast, the phase labelled B preserves time-reversal symmetry, with a sample order parameter being $\Vec{d} = (D_1,D_2,0)^{\mathrm{T}}$, reminiscent of the B phase of $^3$He (with $\lvert \Delta_{\uparrow \uparrow} \rvert = \lvert \Delta_{\downarrow \downarrow} \rvert$). Finally, the phase labelled D is chiral and has a sample order parameter of the form $\Vec{d} = (0,0,D_1 + iD_2)^{\mathrm{T}}$, with a well-known example being the   $p_x + ip_y$ phase (the A phase) of $^3$He.

%
\section{Specific heat of two-component spin-triplet transitions}
\label{sec:SpecificHeat}
%
Here we turn to a discussion of the thermodynamic transitions of two-component spin-triplet superconducting orders, with focus on the entropy and specific heat behavior near the two transition temperatures.

\subsection{General theory}
\label{sec:GeneralTheory}
We consider a single-band superconductor described by the following Bogoliubov-de Gennes (BdG) Hamiltonian at the mean-field level
\begin{equation}
    H_{\mathrm{BdG}} = \frac12 \sum_{\bo{k}} \Vec{\Psi}_{\bo{k}}^{\dagger} \pazocal{H}(\bo{k}) \Vec{\Psi}_{\bo{k}}^{\phantom{\dagger}},
    \label{eq:BdGHam}
\end{equation}
where 
\begin{equation}
    \pazocal{H}(\bo{k}) = \begin{bmatrix}
    \xi(\bo{k})\id & \bo{\Delta}(\bo{k}) \\
    \bo{\Delta}^{\dagger}(\bo{k}) & -\xi(-\bo{k})\id
    \end{bmatrix},
    \label{eq:Hamk}
\end{equation}
in the basis $\Vec{\Psi}_{\bo{k}} = (c_{\bo{k}\uparrow}^{\phantom{\dagger}}, c_{\bo{k}\downarrow}^{\phantom{\dagger}}, c_{-\bo{k}\uparrow}^{\dagger}, c_{-\bo{k}\downarrow}^{\dagger})^{\mathrm{T}}$. We assume inversion symmetry in the normal state, $\xi(\bo{k}) = -\xi(-\bo{k})$, and will henceforth refer to the components of the order parameter in the Balian--Werthamer basis of Eq.~\eqref{eq:OrderParameter}. We consider an order parameter of the form $\bo{\Delta} = (\vec{d} \cdot \vec{\sigma} ) i \sigma_2$ with
\begin{equation}
\begin{aligned}
\vec{d}(T, \bo{k}) &= \Delta_{0} \left[ \sqrt{ 1- \frac{T}{T_{c1}} } \vec{d}_{1}(\bo{k})  + i \varepsilon \sqrt{ 1 - \frac{T}{T_{c2}} } \vec{d}_{2}(\bo{k})  \right],
\end{aligned}
\label{eq:AnsatzGeneral}
\end{equation}
where $\vec{d}_1, \vec{d}_2 \in \mathbb{R}^3$ do not depend on $T$ and belong to distinct odd-parity, 1D irreps of the relevant crystal point group, and where $\varepsilon$ is assumed to be a real parameter controlling the relative size of the components and the strength of TRSB. In general, this ansatz has restricted us to the exotic yet interesting case of a (unitary) triplet order parameter on-setting at $T_{c1}$ with a subsequent second order transition to a composite non-unitary triplet order at $T_{c2}$. 

The specific heat, $C(T) = T \frac{\partial S}{\partial T}$, is derived from the entropy of a Fermi gas
\begin{equation}
\begin{aligned}
 S(T) &= - k_B \sum_{\bo{k}, \sigma} \Big\lbrace f[E_{\sigma}(T,\bo{k})] \ln f[E_{\sigma}(T,\bo{k})] \\
 &\hspace{10pt}+\big(1-f[E_{\sigma}(T,\bo{k})] \big) \ln\big( 1 - f[E_{\sigma}(T,\bo{k})] \big) \Big\rbrace,
\end{aligned}
\label{eq:Entropy}
\end{equation}
where $f(E) = (1+\exp(\beta E))^{-1}$ is the Fermi function, and $E_{\sigma}(T,\bo{k}) = \sqrt{ \xi(\bo{k})^2 + \lvert \Delta_{\sigma}(T,\bo{k}) \rvert^2 }$ are the quasiparticle excitation energies, where $\sigma = \pm$ is indexing the spin, and the spin dependent gaps are given by Eq.~\eqref{eq:SpinSplitGaps} with a mean-field temperature dependence. We invoke two standard assumptions when calculating the heat capacity from Eq.~\eqref{eq:Entropy}. First, the momentum sum is replaced by integrals over $(\xi, \bo{k})$ where now $\bo{k}$ lies on the iso-surface $\xi(\bo{k}) = \xi$: $\sum_{\bo{k}} \cdot \to \int_{-\omega_c}^{\omega_c} \D \xi~\langle \cdot \rangle_{\mathrm{FS}}$, where $\omega_c$ is the cutoff (e.g.,~the electronic bandwidth), $\langle A \rangle_{\mathrm{FS}} = \int_{S_{\mathrm{F}}} \frac{\D \bo{k}}{(2\pi)^3} \frac{A_{\sigma}}{v_{\mathrm{F}}(\bo{k})}$, and $v_{\mathrm{F}}(\bo{k}) = \lvert \nabla_{\bo{k}} \xi(\bo{k}) \rvert$ is the Fermi velocity. This approximation is justified in the thermodynamic limit. Second, we assume weak coupling, $\lvert \Delta \rvert,~T \ll \omega_c$ such that the $\xi$ integration limits can be extended~\footnote{The integrand in the heat capacity has support over a range controlled by $T$ in the weak-coupling limit since $\xi^2 / [4\cosh^2(\xi/[2k_B T])] \approx \xi^2 \exp(-\xi/[k_B T])$, and the standard deviation of the latter distribution is $\sqrt{3} k_B T$.}. The heat capacity becomes:
\begin{align}
C(T) &= \f{1}{k_B T^2} \int_{-\infty}^{\infty} \D \xi \hspace{1mm} \sum_{\sigma} \Big\langle \f{\xi^2 + \lvert \Delta_{\sigma} \rvert^2 - \f{T}{2} \f{\partial \lvert \Delta_{\sigma} \rvert^2}{\partial T} }{ 4 \cosh^2(\f{E_{\sigma}}{2k_B T})} \Big\rangle_{\mathrm{FS}}.
\label{eq:Specificheat}
\end{align}
The term containing the temperature derivative of the gaps in Eq.~\eqref{eq:Specificheat} is responsible for a discontinuous jump in $C(T)$ at the onset of the order parameter. In the scenario of two symmetry-distinct order parameter components as considered in the preceding section, discontinuous jumps occur at both $T_{c1}$ and $T_{c2}$. Focusing on the second onset ($T_{c2}$), which in the A phase of the preceding section marks the transition from a unitary to a non-unitary state, the specific heat capacity anomaly is quantified by the difference:
\begin{equation}
\Delta C(T_{c2}) \equiv C(T_{c2}^-) - C(T_{c2}^+) = \f{1}{8k_B T_{c2}} \int_{-\infty}^{\infty} \D \xi~\delta c, 
\label{eq:HeatCapacityMassage1}
\end{equation}
where $T = T_{c2}^{\pm}$ refers to taking the one-sided limits $\lim_{T\to T_{c2}^{\pm}}$, approaching $T_{c2}$ from above and below, and where
\begin{equation}
\begin{aligned}
\delta c &= \sum_{\sigma} \Big\langle \Big[ \f{\partial \lvert \Delta_{\sigma}(T,\bo{k})\rvert^2}{\partial T}\Big\rvert_{T_{c2}^+} - \f{\partial \lvert \Delta_{\sigma}(T,\bo{k}) \rvert^2}{\partial T}\Big\rvert_{T_{c2}^-}  \Big] \\
&\hspace{10pt}\times\sech^{2}\Big( \f{E_{\sigma}(T_{c2},\bo{k})}{2k_B T_{c2}} \Big) \Big\rangle_{\mathrm{FS}}.
\end{aligned}
\label{eq:HeatCapacityMassage}
\end{equation}
We use the order parameter ansatz of Eq.~\eqref{eq:AnsatzGeneral} in Eq.~\eqref{eq:HeatCapacityMassage} and obtain two contributions. The first contribution comes from $\frac{\partial \lvert \vec{d} \rvert^2}{\partial T}$ and is \emph{positive semidefinite}, hence supplying $\Delta C$ with an anticipated positive semidefinite contribution, similar to that reported in the spin-singlet scenario of Ref.~\cite{Kivelson2020}. A second and non-standard contribution, however, comes from $-\frac{\partial \lvert \vec{d}^{\ast} \times \vec{d} \rvert }{\partial T}\big\rvert_{T_{c2}^-}$ due to the spin-split gaps for non-unitary orders in Eq.~\eqref{eq:SpinSplitGaps}. The latter contribution is explicitly \emph{negative semidefinite} because $E_{-}(T_{c2}^{-},\bo{k}) \leqslant E_{+}(T_{c2}^{-},\bo{k})$ for all crystal momenta, and it is only finite for non-unitary states. This gives rise to the following exact result including both terms discussed above:
\begin{widetext}
\begin{equation}
\delta c = \frac{2 (\varepsilon \Delta_0)^2}{T_{c2}} \Big\langle \sech^{2}\Big( \frac{\sqrt{\xi^2+g^2} }{2k_B T_{c2}} \Big) \Big[  \lvert \Vec{d}_2 \rvert^2 - \frac{\Delta_0^2}{k_B T_{c2}} \frac{\lvert \vec{d}_1 \times \vec{d}_2 \rvert^2 }{\sqrt{\xi^2 + g^2}} \left( 1-\frac{T_{c2}}{T_{c1}} \right) \tanh \Big( \frac{\sqrt{\xi^2+g^2} }{2k_B T_{c2}} \Big) \Big] \Big\rangle_{\mathrm{FS}},
\label{eq:Jump}
\end{equation}
\end{widetext}
where
\begin{equation}
g \equiv \Delta_0 \lvert \vec{d}_1 \rvert \sqrt{1 - \frac{T_{c2}}{T_{c1}}}.
\label{eq:gDef}
\end{equation}
This result shows how the non-unitarity of the order parameter is associated with a \emph{negative} contribution to the specific heat discontinuity that can result in a partly or entirely suppressed, or even net negative, secondary specific heat jump. We note that the formula above contains a Fermi surface average as also discussed in view of ferromagnetic and antiferromagnetic non-unitary pairing states~\cite{Volovik1985}, but the additional non-constant terms that are multiplied before the average and the fact that the square $\lvert \vec{d}_1 \times \vec{d}_2 \rvert^2$ enters, disallows a direct connection. Further, we note that ferromagnetic pairing states can generally have a larger contribution.

From the entropic point of view, the sign of the second heat capacity anomaly (Eq.~\eqref{eq:HeatCapacityMassage}) is simply related to the ``sign'' of the non-analyticity of $S$, i.e., $\mathrm{sign}[\frac{\partial S}{\partial T}\big\rvert_{T_{c2}^-} - \frac{\partial S}{\partial T}\big\rvert_{T_{c2}^+}]$. This is further explained in Appendix~\ref{app:Entropy}. Though we have not proven the stability of the order parameter considered, there is nothing at the thermodynamic level that formally disallows the unusual negative sign of the heat capacity discontinuity. In Appendix~\ref{app:Entropy} we also discuss how the above result generalizes to critical exponents beyond mean field. Moreover, while intraband couplings can affect the power law of the temperature profile of the gap near the lower transition~\cite{SuhlMatthiasWalker59}, we stress that our ansatz in Eq.~\eqref{eq:AnsatzGeneral} concerns two symmetry-distinct second-order transitions, for which coupling terms at quartic order in the free energy can lead to a renormalization of $T_{c2}$ while leaving the critical exponent unaltered~\cite{Roising2022}.

One may question whether a single-component non-unitary order is sufficient to obtain the anomalous heat capacity behaviour above. This case is considered in Appendix~\ref{app:SingleNonUnitary} and turns out to always have a positive heat capacity jump. Technically, this is because a diverging $\partial \lvert \vec{d}^{\ast} \cross \vec{d} \rvert / (\partial T)$ at $T = T_{c2}^-$ is needed to give a finite-valued outcome when multiplied with $\sum_{\sigma} \sigma\sech^2(E_{\sigma}/(2k_B T_{c2}^-))$, which approaches zero as $\sqrt{\delta T}$ when expressing $T_{c2}^- = T_{c2}-\delta T$ in the two-component case. In the single-component scenario, the prior factor does not diverge, which emphasizes that the negative heat capacity contribution hinges on a transition splitting, $T_{c2} < T_{c1}$, as also reflected in the negative term being proportional to $1-T_{c2}/T_{c1}$ in Eq.~\eqref{eq:Jump}.

Another generic observation from Eq.~\eqref{eq:Jump} that impacts the negative contribution can be pointed out. The negative term has a prefactor of $\Delta_0/(k_B T_{c2})$ which in BCS theory takes the conventional value of $\pi e^{-\gamma} \approx 1.764$. It is well known that both gap anisotropies (at weak coupling)~\cite{Einzel03}, as well as strong-coupling effects~\cite{Carbotte90} can increase this ratio, both of which enhance the unusual negative jump effect.

%
\section{Application to $\mathrm{UTe}_2$}
\label{sec:UTe2}
%
In this section we perform a material-specific study of two-component spin-triplet superconductivity applied to UTe$_2$. In this compound non-unitary superconducting states from different irreps are actively discussed as candidate states for explaining several experimental findings such as TRSB~\cite{HayesEA21, Ajeesh2023} and chiral edge modes~\cite{Jiao2020}. We start the section with a detailed discussion of the electronic structure of UTe$_2$. This allows us to discuss thermodynamic transitions of UTe$_2$ and illustrate the unusual specific heat behavior that may be associated to two-component non-unitary spin-triplet superconductivity. Finally we present a material-specific discussion of the topological properties of superconducting UTe$_2$.

\subsection{Electronic structure}
\label{sec:electronicUTe2}
\begin{figure}[tb]
	\centering
	\includegraphics[width=\linewidth]{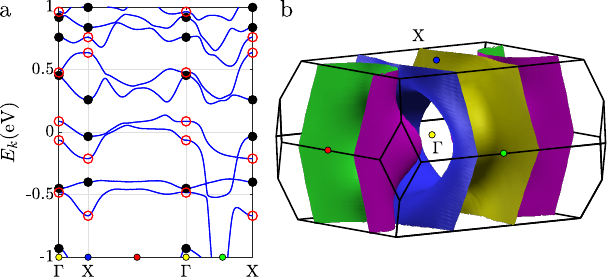}
    \caption{(a) Bandstructure along the three principal axis where the BZ boundaries in the $\hat{x}$, $\hat{y}$ and $\hat{z}$ directions are marked by red, green, and blue dots, respectively. The symmetry of the bands under inversion operation are marked by red circles ($+1$) and full black dots ($-1$). (b) Fermi surface as obtained from an {\it ab initio} calculation for $U=1.2\;\mathrm{eV}$ and plotted with software described in Ref.~\cite{Kokalj1999}.}
	\label{fig:bands_FS}
\end{figure}

Experimentally, the specific heat was measured on clean samples of UTe$_2$ exhibiting
higher $T_c$'s of around $2$~K and large residual resistivity ratios of several hundreds, finding: (i) a single specific heat capacity transition and (ii), a $C/T \sim T^2$ tail consistent with point nodes in the superconducting gap~\cite{Rosa2022,Cairns2020,Girod2022,AokiEA22}. For quantitative calculations of the specific heat and the detailed functional form of $C/T$, a precise description of (a) the superconducting order parameter and (b), the low-energy electronic structure including the Fermi surface shape and the Fermi velocities are needed. Given the magnetic susceptibility at low temperature~\cite{Ran2019} compatible with a Fermi liquid, we assume a picture of itinerant electrons where U 5$f$-states contribute.

As earlier works~\cite{Xu2019,IshizukaEA19,ShishidouEA21,Kreisel2022,EatonEA23}, we adopt the approach of considering a series of electronic structures from a DFT+$U$ calculation where the effective $U$ is a free parameter eventually fixed by comparison to spectroscopic data, and discuss common properties of the low-energy electronic structure and the symmetry-based indicators in Sec.~\ref{sec:UTe2_topology_SI}. The starting point is the body-centered orthorhombic lattice structure of UTe$_2$ with space group $Immm$ and lattice structure as determined experimentally~\cite{Ikeda2006}. We use the WIEN2K package~\cite{Blaha2020} with the generalized gradient approximation [Perdew-Burke-Ernzerhof (PBE) functional]~\cite{Perdew1996}, use a k-mesh of $5000~\bo{k}$ points ($17^3$) together with with RMT $\times$ KMAX of $9.0$ in a relativistic calculation including SOC on all atoms and adding correlations on the U 6d and 5f electrons with the parameter $U$ while keeping the Hund's exchange interaction $J=0$. In this setting, we obtain an insulating state at $U=0$ which becomes metallic at $U_0=0.97\;\mathrm{eV}$ with a band of mixed parity crossing the Fermi level between $\Gamma$ and $\mathrm{X}$~\cite{ShishidouEA21} and a second corrugated cylindrical Fermi pocket which grows and eventually vertically spans the Brillouin zone boundary. Increasing further to $U_1=1.03\;\mathrm{eV}$, a small pocket at $\Gamma$ appears which quickly is pushed down at $U_3=1.06\;\mathrm{eV}$ again to yield a Fermi surface topology with the important band crossing between $\Gamma$ and $\mathrm{X}$ present over a sizeable range of $U$, see Fig.~\ref{fig:bands_FS}. At $U_4=1.44\;\mathrm{eV}$ this band crossing is lifted and the electronic structure resembles the one of a putative ThTe$_2$ calculation~\cite{MiaoEA20,Kreisel2022} with slightly corrugated Fermi surfaces; increasing the correlations further, reduces the corrugation and increases the Fermi velocities. Calculations with finite Hund's interaction $J=0.1U$ and $J=0.2U$ give qualitatively similar results with the energy scales $U_i$ shifted upwards. Having these quantitative and qualitative variations in mind, one can try to pinpoint the relevant regime by comparing to ARPES and quantum oscillation data to determine the topology of the Fermi surface.

The ARPES data in Ref.~\cite{Fujimori19} have been interpreted as existence of a hole pocket around the $\Gamma$ point, while another work found a dispersive band dropping down at the ``Z point''~\cite{Shick21} which is labeled X in our notation. Reference~\cite{MiaoEA20} shows data consistent with two-dimensional (2D) cylindrical tubes of Fermi surfaces. More recent experiments on cleaner samples detect a cylindrical-shaped electron Fermi surface without connections at the ``Z point'' of the Brillouin zone (BZ)~\cite{Aoki23}, in agreement with the 3D conductivity component from resistivity measurements together with an analysis of scattering rates as detected by ARPES~\cite{Eo22}. Recently, quantum oscillation measurements in clean crystals reported the finding of several frequencies consistent with 2D Fermi surfaces with little corrugation~\cite{AokiEA22}. Another work observed a low-frequency component reminiscent of a 3D Fermi surface pocket from a band with moderately small effective mass  $m_*=5.7m_e$~\cite{BroylesEA23}. This is in contradiction to a mapping of the Fermi surface from quantum oscillations finding only 2D Fermi surfaces and constraining any 3D Fermi surface to a very small volume or exhibiting extremely large effective masses $m_* > 78m_e$ to render it unobservable at the base temperature of the experiment~\cite{EatonEA23}. 

In summary, there is experimental evidence for a Fermi surface of UTe$_2$ similar both to the intermediate-$U$ and the large-$U$ regime of the DFT+$U$ calculations. In the following we pursue mainly the intermediate-$U$ case with the Fermi surface topology shown in Fig.~\ref{fig:bands_FS}. For the strong topological phases discussed in this work, the Fermi surface is equivalent to 2D sheets and a closed pocket around X.

\subsection{Thermodynamic transitions and specific heat}
\label{sec:specificUTe2}
As evident from the above, both the Fermi surface shape and topology and the detailed spin- and momentum-dependent structure of the superconducting order parameter of UTe$_2$ are currently matters of substantial controversy. Therefore, we restrict the study of the specific heat to a qualitative analysis, and return to a discussion of the consequences of the detailed electronic structure of UTe$_2$ in the topology section. To examine the effects of non-unitary pairing states on-setting at a second triplet order parameter transition, we simply use a model of a quadratic bandstructure and select among possible triplet superconducting order parameters for $D_{2\mathrm{h}}$ and leave any quantitative calculation of the specific heat to future studies once the Fermi surface of UTe$_2$ and the superconducting order are better determined.
 
\begin{figure}[bt]
	\centering
	\includegraphics[width=\linewidth]{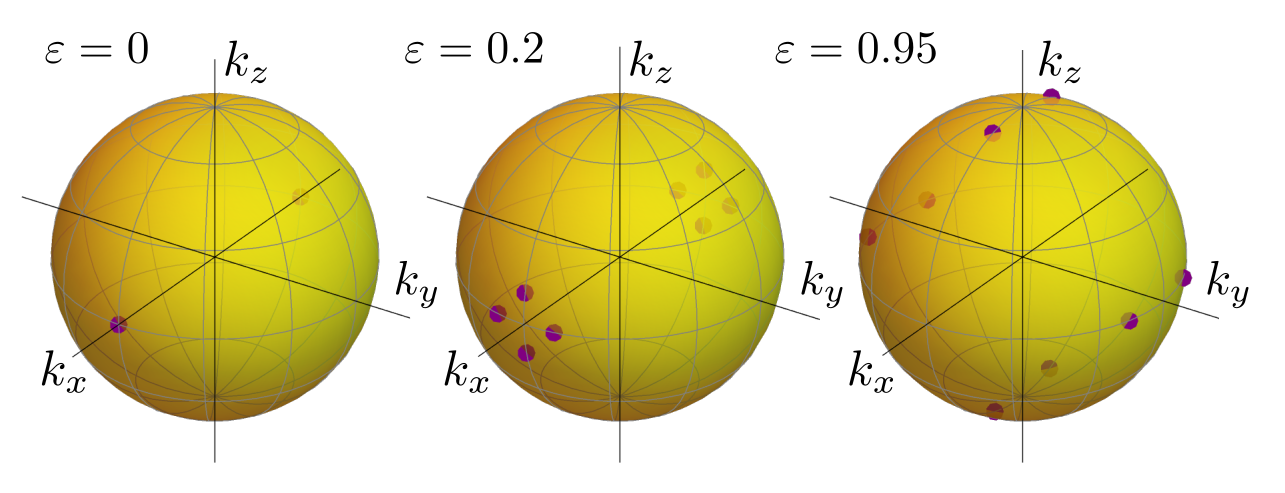}
    \caption{Nodes of the zero-temperature gap $\lvert \Delta_{\sigma=-} \rvert$, as obtained from Eq.~\eqref{eq:SpinSplitGaps} for the order parameter $\Vec{d}_{\mathrm{B}_{3\mathrm{u}}} + i \varepsilon \Vec{d}_{\mathrm{A_u}}$, cf.~Tab.~\ref{tab:IrrepsD2hSOC}. Locations of point nodes are indicated with purple dots. This scenario was studied in the context of penetration depth measurements of UTe$_2$ in Ref.~\cite{KotaEA23}. For $\lvert \varepsilon \rvert > 1$ the state is fully gapped.}
	\label{fig:Gaps}
\end{figure}

\begin{figure*}[t!bh]
	\centering
	\includegraphics[width=\linewidth]{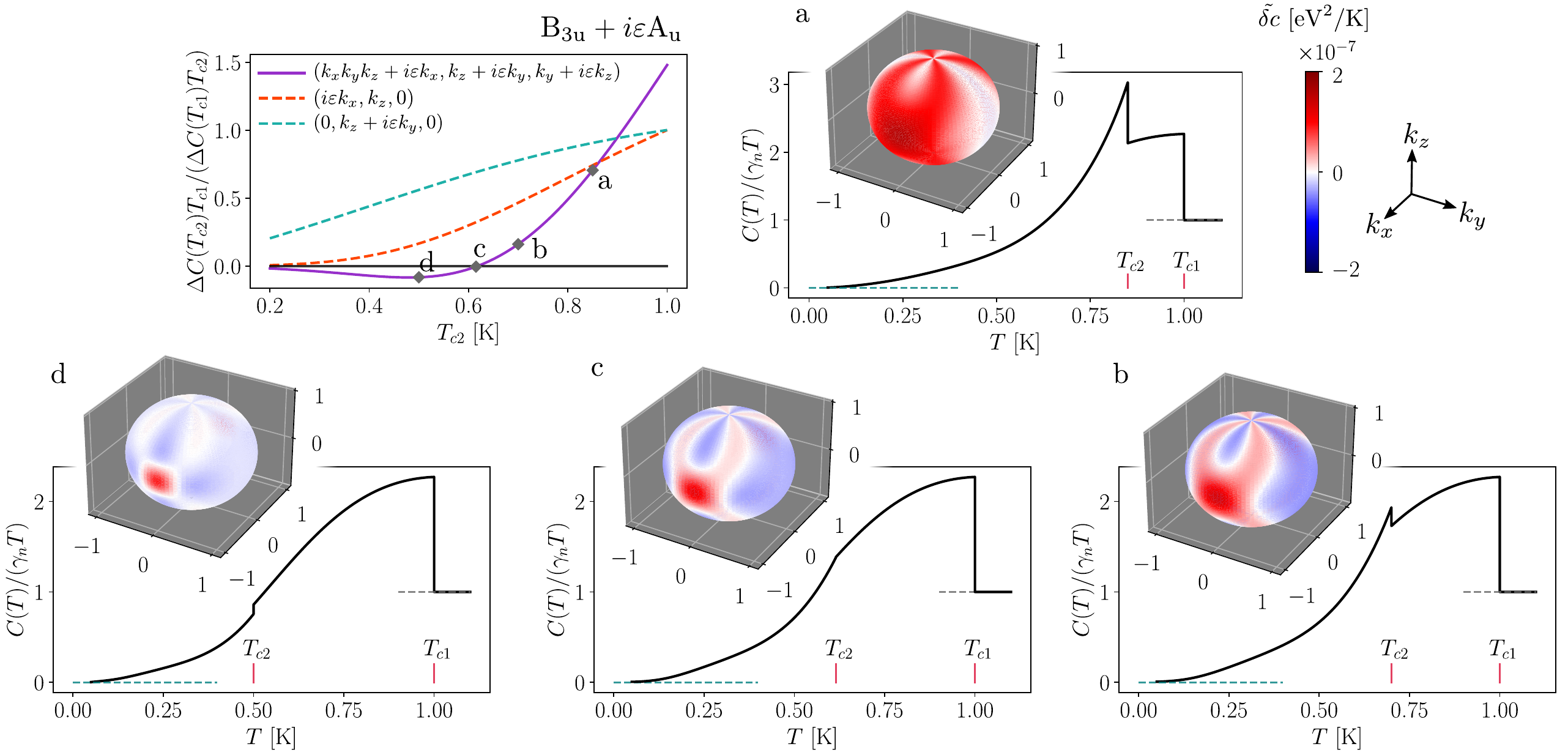}
    \caption{Specific heat jumps for the triplet order parameter $\vec{d}(T,\bo{k})$ of Eq.~\eqref{eq:AnsatzGeneral} for irreps $\mathrm{B}_{3\mathrm{u}}$ (on-setting at $T_{c1}$) and $\mathrm{A}_{\mathrm{u}}$ (on-setting at $T_{c2}$) with an orthorhombic crystal (point group $D_{2\mathrm{h}}$) and a spherical Fermi surface for several sets of coefficients $c$ as explained in the labels, cf.~Tab.~\ref{tab:IrrepsD2hSOC}. Panels (a)-(d) show the specific heat per temperature for the case shown in purple in the top left panel. The insets, with colorscale and momentum axes defined to the right of panel (a), display the integrand (at $\xi = 0$) of Eq.~\eqref{eq:Jump}, labelled $\tilde{\delta c}$, resolved over the Fermi surface to reveal positive and negative contributions to the secondary specific heat jump. Panel (c) is the fine-tuned case of $T_{c2} = 0.615 T_{c1}$ at which the second heat capacity jump vanishes and the phase transition is effectively third order. In these calculations we kept $\varepsilon = T_{c2}/T_{c1}$ and a mean-field gap value of $\Delta_0 = 2 \times 1.764 k_B T_{c1}$.}
	\label{fig:HeatCapacity}
\end{figure*}

In terms of possible TRSB pairing candidates for UTe$_2$, motivated by recent experimental developments~\cite{KotaEA23} we consider initially the case of $\vec{d}_1 = \vec{d}_{\mathrm{B}_{3\mathrm{u}}}$ and $\vec{d}_2 = \vec{d}_{\mathrm{A}_{\mathrm{u}}}$ belonging to point group $D_{2\mathrm{h}}$ in the presence of spin-orbit coupling (Tab.~\ref{tab:IrrepsD2hSOC}). Locations of the point nodes of the nodal gap in the spherical Fermi surface are shown in Fig.~\ref{fig:Gaps} for different values of the ``mixing parameter'' $\varepsilon$. Calculations of $C(T)/T$ per normal state value from Eq.~\eqref{eq:Specificheat} are shown in Fig.~\ref{fig:HeatCapacity}. The insets of Fig.~\ref{fig:HeatCapacity} display the integrand of Eq.~\eqref{eq:Jump}, i.e.,~the value of the quantity before integration over the Fermi surface, at $\xi = 0$. Clearly, the momentum structure of this quantity is dictated by two competing terms of Eq.~\eqref{eq:Jump}, involving both $\lvert \vec{d}_1(\bo{k}) \rvert$, $\lvert \vec{d}_2(\bo{k}) \rvert$, and $\lvert \vec{d}_1(\bo{k}) \times \vec{d}_2(\bo{k}) \rvert$ (see also Appendix~\ref{app:Entropy}). As seen directly from Eq.~\eqref{eq:Jump} and~\eqref{eq:gDef}, reducing $T_{c2}$ while keeping $T_{c1}$ fixed is identified as an efficient way to increase the relative impact of the negative contribution to $\Delta C$. 
\begin{table}[b]
\caption{Odd-parity irreducible representations of $D_{2\mathrm{h}}$ (including SOC~\cite{Annett90}). In the second column $X$ represents any function that transforms like $\sin{k_x}$ under the point group operations and similar for $Y$ and $Z$. In the third column: ``p.'' refers to point nodes with the locations on a spherical Fermi surface indicated in a parenthesis~\cite{AokiEA22_rev}. The coefficients $c_1$, $c_2$, $c_3$ are real, but otherwise unrestricted by the point group.}
\begin{center}
\begin{tabular}{p{1.0cm} p{4.0cm} p{1.8cm}}
\toprule
 Irrep. & Order parameter & Nodes \\ \midrule
 $\mathrm{A}_{\mathrm{u}}$ & $\vec{d} = (c_1 X, c_2 Y, c_3 Z)^{\mathrm{T}}$ & gapped  \\
 $\mathrm{B}_{1\mathrm{u}}$ & $\vec{d} = (c_1 Y, c_2 X, c_3 XYZ)^{\mathrm{T}}$ & p.~(along $\hat{z}$) \\
 $\mathrm{B}_{2\mathrm{u}}$ & $\vec{d} =  (c_1 Z, c_2 XYZ, c_3 X )^{\mathrm{T}}$ & p.~(along $\hat{y}$) \\
 $\mathrm{B}_{3\mathrm{u}}$ & $\vec{d} = (c_1 XYZ, c_2 Z, c_3 Y)^{\mathrm{T}}$ & p.~(along $\hat{x}$) \\
  \bottomrule
\end{tabular}
\end{center}
\label{tab:IrrepsD2hSOC}
\end{table}

As also demonstrated by Fig.~\ref{fig:HeatCapacity}(d) it is possible to obtain a specific heat drop at the second transition, signifying a slower entropy decrease with decreasing temperatures at $T$ lower than $T_{c2}$. Although generically requiring fine tuning,  a scenario in which the positive and negative terms cancel to yield a vanishing second heat capacity anomaly is also conceivable (Fig.~\ref{fig:HeatCapacity}(c)), effectively making the transition third order. For the parameters used in Fig.~\ref{fig:HeatCapacity} this happens around $T_{c2} = 0.615 T_{c1}$. Smearing from spatial inhomogeneities may further wash out any signature of the second transition~\cite{Andersen2006T,Roising2022}.

Due to the assumptions of the spherical Fermi surface, the results for $C(T)/T$ in Fig.~\ref{fig:HeatCapacity} remain unchanged had we instead used $\vec{d}_1 = \vec{d}_{\mathrm{B}_{1\mathrm{u}}}$ or $\vec{d}_1 = \vec{d}_{\mathrm{B}_{2\mathrm{u}}}$. In addition, the important ingredient for acquiring an anomalous contribution to the specific heat is the non-unitary nature of the two-component $d$-vector. Therefore, pairing states of the form $\mathrm{B}_{1\mathrm{u}}+i\varepsilon \mathrm{B}_{2\mathrm{u}}$ and similar combinations also exhibit such unusual thermodynamic transitions, including the possibility of a vanishing second specific heat anomaly. These states turn out to be of the type ferromagnetic non-unitary pairing states~\cite{Volovik1985} unless fine-tuned by parameters to yield (accidental) vanishing magnetization.

\subsection{Topological properties}
\label{sec:TopPhases}
In this section, we begin by determining the topological phases and corresponding anomalous boundary states of a Bogoliubov-de Gennes (BdG) Hamiltonian describing a spherical Fermi surface with $D_{2\mathrm{h}}$ point group symmetry and various superconducting orders. We discuss all pure odd-parity orders as well as mixtures where one order is considered as an infinitesimal perturbation to the other. Here, we focus on the anomalous boundary phenomenology of the topological phases and leave the validation of the topological phases to Appendix~\ref{app:topo_validation}.
Table~\ref{tab:topological_phases} summarizes the topological phases obtained for the considered superconducting orders. 

Next, in section~\ref{sec:UTe2_topology_SI}, we apply the theory of symmetry-based indicators to predict the topology of the superconducting phases obtained by including superconducting pairing in the bandstructure calculated from DFT+$U$ as summarized in Sec.~\ref{sec:electronicUTe2}. These results support that the topological phenomenology described for a spherical Fermi surface may indeed apply to ${\rm UTe}_2$.

\subsubsection{Nodal phases}
\label{sec:TopPhases_nodal}
The nodal points in the quasiparticle spectrum at the intersection of the Fermi surface with the nodal lines of the order parameters are protected by a topological invariant defined on an enclosing surface of the reciprocal-space BdG Hamiltonian. By the bulk-boundary correspondence, this topological invariant has associated anomalous boundary excitations \cite{SimonPRB2022} that we describe in the following. 

\begin{figure}[tb]
    \centering
    \includegraphics[width=\columnwidth]{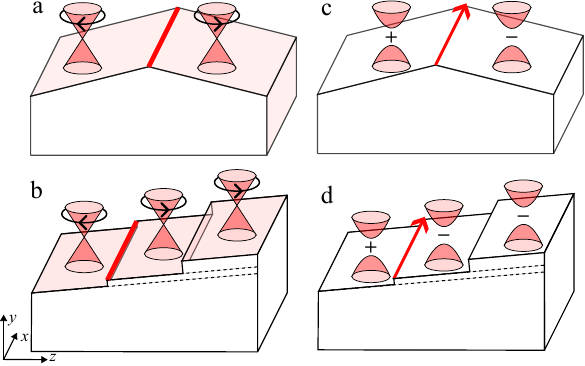}
    \caption{Sketch of the anomalous boundary signatures around $k_x = 0$ of the nodal topological phase of the spherical Fermi pocket with (a), (b) ${\rm B}_{3\mathrm{u}}$ order parameter and (c), (d) ${\rm B}_{3\mathrm{u}} + i \varepsilon \mathrm{B_{2u}}$ order parameter symmetry. The top row displays a mirror $z \to -z$ symmetric hinge. The bottom row displays step edges on an asymmetric hinge. With a ${\rm B}_{3\mathrm{u}}$ order parameter [(a), (b)], surfaces that respect $k_x$ translation symmetry have a Majorana Dirac surface cone. Dirac cones on surfaces with opposite chirality host a flat Majorana band at the interface (thick red line). For the mixed order parameter with ${\rm B}_{3\mathrm{u}} + i \varepsilon \mathrm{B_{2u}}$ symmetry [(c), (d)], the Majorana Dirac surface cones gap out due to breaking of time-reversal symmetry. Interfaces between surfaces with opposite sign of the mass term host a chiral Majorana mode (thick red arrow).}
    \label{fig:sketch_topo}
\end{figure}
\begin{table*}
\caption{Summary of the topological phases obtained for a single, spin-degenerate band with spherical
Fermi surface and $D_{2\mathrm{h}}$ point group symmetry and superconducting
order with symmetry specified by the irreducible representations in the
left column. 
For the mixed orders, the factor $\lvert \varepsilon \rvert \ll 1$ indicates that the corresponding order is an infinitesimal perturbation to the other.  
The right column specifies possible gapless excitations on a $(0, -1, 1)$ crystal surface that are associated to the bulk topology, in addition to bulk nodes.  Further details on each phase are given in the main text. $(\dagger)$ The entries for the mixture $\mathrm{B}_{3\mathrm{u}}+i\mathrm{A_{u}}$ considers a specific choice of a chiral order parameter $\vec{d}(\bo{k}) = (0,c_{1}k_{z}+ic_{2}k_{y},0)^{{\rm T}}$, and similarly for the mixtures $\mathrm{B}_{1\mathrm{u}}+i\mathrm{A_{u}}$ and $\mathrm{B}_{2\mathrm{u}}+i\mathrm{A_{u}}$. 
}
\begin{tabular}{p{3.6cm}p{6.5cm}p{7cm}}
\toprule
Irrep. & Topological phase & $(0,-1,1)$ surface\tabularnewline \midrule
$\mathrm{B_{1u}}$, $\mathrm{B_{2u}}$, $\mathrm{B_{3u}}$  & \emph{Second-order topological nodal superconductor of type (ii)}~\cite{SimonPRB2022} & Majorana Dirac cone protected by time-reversal symmetry and bulk nodes\tabularnewline
$\mathrm{B_{3u}}+i\varepsilon\mathrm{B}_{2u}$, $\mathrm{B_{2u}}+i\varepsilon\mathrm{B}_{3u}$,  
$\mathrm{B_{3u}}+i\varepsilon\mathrm{B}_{1u}$, $\mathrm{B_{1u}}+i\varepsilon\mathrm{B}_{3u}$  & \emph{Second-order topological nodal superconductor of type (iii)}~\cite{SimonPRB2022} & Fermi arcs of Bogoliubov quasiparticles around the projection of the Weyl nodes  \tabularnewline  \tabularnewline
$\mathrm{B_{2u}}+i\varepsilon\mathrm{B}_{1u}$, $\mathrm{B_{1u}}+i\varepsilon\mathrm{B}_{2u}$ & \emph{Second-order topological nodal superconductor of type (iii)}~\cite{SimonPRB2022} & Majorana Dirac cone protected by mirror symmetry and Fermi arcs of Bogoliubov quasiparticles around the projection of the Weyl nodes \tabularnewline
$\mathrm{B_{3u}}+i\varepsilon\mathrm{A_{u}}$ & \emph{Second-order topological superconductor}~\cite{GeierPhysRevB2018May} \emph{with coexisting Weyl
nodes} & Majorana Dirac cone protected by mirror symmetry and Fermi arcs of Bogoliubov quasiparticles around the projection of the Weyl nodes \tabularnewline
$\mathrm{B_{1u}}+i\varepsilon\mathrm{A_{u}}$, $\mathrm{B_{2u}}+i\varepsilon\mathrm{A_{u}}$ & \emph{Second-order topological superconductor}~\cite{GeierPhysRevB2018May} \emph{with coexisting Weyl
nodes} & Fermi arcs of Bogoliubov quasiparticles around the projection of the Weyl nodes \tabularnewline
$\mathrm{B_{1u}}+i\mathrm{A_{u}}$, $\mathrm{B_{2u}}+i\mathrm{A_{u}}$,
$\mathrm{B_{3u}}+i\mathrm{A_{u}}$ $(\dagger)$ & \emph{Weyl superconductor}~\cite{MengPhysRevB2012Aug} & Large Fermi arcs of Bogoliubov quasiparticles connecting Weyl nodes with positive and negative $k_x$ \tabularnewline \tabularnewline
$\mathrm{A_{u}}$ & \emph{First-order topological superconductor}~\cite{ChiuEA16} & Majorana Dirac cone protected by time-reversal symmetry\tabularnewline
$\mathrm{A_{u}}+i\varepsilon\mathrm{B_{3u}}$ & \emph{Second-order topological superconductor}~\cite{GeierPhysRevB2018May} & Majorana Dirac cone protected by mirror symmetry \tabularnewline
$\mathrm{A_{u}}+i\varepsilon\mathrm{B_{1u}}$, $\mathrm{A_{u}}+i\varepsilon\mathrm{B_{2u}}$ & \emph{Second-order topological superconductor}~\cite{GeierPhysRevB2018May} & Gapped \tabularnewline \bottomrule
\end{tabular}
\label{tab:topological_phases}
\end{table*}

{\em ${\rm B}_{3\mathrm{u}}$ superconducting order.---} A superconducting order parameter with ${\rm B}_{3\mathrm{u}}$ symmetry has a nodal line along the $k_x$-axis that intersects with a spherical Fermi surface at $(k_{\rm F},0,0)^{\rm T}$ as shown in the leftmost panel of Fig.~\ref{fig:Gaps}. As a function of $k_x$, the node coincides with a change of a second-order topological invariant protected by mirror symmetry and chiral antisymmetry defined on 2D slices with fixed $k_x$ within the 3D BZ (see Appendix~\ref{app:topo_validation} for a validation of the result). The order of the topological phase determines the dimensionality of the corresponding anomalous surface states \cite{SchindlerSciAdv2018Jun, GeierPhysRevB2018May}. At $k_x = 0$, time-reversal symmetry requires that this 2D topological phase is first order. \phantom{\cite{FuPhysRevLett2007Mar}}

Altogether, the boundary signatures can be understood as follows. As sketched in Fig.~\ref{fig:sketch_topo}(a), mirror-symmetry breaking surfaces host a single Majorana Dirac surface cone around $k_x = 0$~\footnote{The attribute ``Majorana'' indicates that the Dirac cone describes Bogoliubov quasiparticles that are their own antiparticle~\cite{ReadGreen00, QiPhysRevLett2009May, ChiuEA16}. The Majorana Dirac cone thereby describes a single massless Majorana fermion in two dimensions. This property distinguishes the surface Majorana Dirac cone from surface Dirac cones in topological insulators~\cite{FuPhysRevLett2007Mar} where the quasiparticles are ordinary Dirac fermions.}. Mirror symmetries ${\cal M}_y$ and ${\cal M}_z$ require that surfaces whose orientation is related by mirror symmetry host Majorana Dirac surfaces cones with opposite chirality. As a consequence, a mirror-symmetric hinge hosts a flat band of zero-energy states. In reciprocal space, the flat band connects to the bulk node at $k_x = k_F$ and disappears thereafter. These are typical boundary signatures of second-order topological nodal superconductors of type (ii)~\cite{SimonPRB2022}.

Step edges on an asymmetric surface, Fig.~\ref{fig:sketch_topo}(b), also host a flat band at zero energy if the chirality of the Majorana Dirac surface cones on adjacent surfaces are opposite. In experiment, this may be the case if the chemical structure of the surfaces are different, for example if the step edge has a fractional unit cell height. If the step edge has a height of one unit cell, the chirality of the Majorana Dirac surface cones is the same unless there is another structural difference between the surfaces.

In ${\rm UTe}_2$, phenomenology consistent with this topological phase has been discussed in Ref.~\cite{IshizukaEA19}: in this reference, a weak topological invariant $\nu_1$ has been calculated for a bandstructure obtained from DFT+$U$ calculations. Together with the point node for a superconducting order $\mathrm{B_{3u}}$ symmetry, this topological invariant $\nu_1$ detects the Majorana Dirac surface cones that are present in this nodal topological superconducting phase. 

We further describe the properties of a $(0, -1 , 1)$ crystal surface relevant to experiments in ${\rm UTe}_2$~\cite{MadhavenEA20}. The surface preserves translation symmetries along the $(1,0,0)$ and $(0,1,1)$ directions as well as mirror symmetry ${\cal M}_x$, but breaks the remaining crystalline symmetries. This surface hosts a Majorana Dirac cone and may host flat zero-energy states on step edges between chemically distinct surfaces as described above. The surface modes hybridize with low-energy bulk modes around the projection of the bulk nodes onto the surface BZ. 

{\em ${\rm B}_{1\mathrm{u}}$ and ${\rm B}_{2\mathrm{u}}$ superconducting order.---} The results for ${\rm B}_{1\mathrm{u}}$ [${\rm B}_{2\mathrm{u}}$] symmetric superconducting order are equivalent to the results of $\mathrm{B_{3u}}$ up to a permutation of coordinates $(x,y,z) \to (z,x,y)$ [$(x,y,z) \to (y,z,x)$]. For ${\rm B}_{1\mathrm{u}}$ [${\rm B}_{2\mathrm{u}}$], the $(0, -1 , 1)$ surface corresponds to the $(1, 0 , -1)$ [$(-1, 1 , 0)$] surface when permuting the coordinates such that the order corresponds to the ${\rm B}_{3\mathrm{u}}$ irrep. In both cases, the surface hosts a Majorana Dirac cone and the bulk nodes project onto the surface BZ at opposite momenta.

Ref.~\cite{TeiEA23} discusses a mirror Chern number as well as winding numbers as topological invariants to characterize a model for ${\rm UTe}_2$ with $\mathrm{B_{1u}}$, $\mathrm{B_{2u}}$, and $\mathrm{B_{3u}}$ pairing. The mirror Chern number also detects a Majorana Dirac surface cone on mirror-symmetric surfaces. The parity of the mirror Chern number is equal to the weak invariant $\nu_j, \ j = 1,2,3$ computed from the normal-state Fermi pockets in the respective plane as analyzed in Ref.~\cite{IshizukaEA19}. 

{\em ${\rm B}_{3\mathrm{u}} + i \varepsilon \mathrm{B_{2u}}$ superconducting order.---} In the presence of an additional, infinitesimal ($|\varepsilon| \ll 1$) superconducting order with $\mathrm{B_{2u}}$ symmetry and a superconducting phase relative to the dominant $\mathrm{B_{3u}}$ symmetric order, time-reversal symmetry as well as two-fold rotation ${\cal R}_{x}$, ${\cal R}_{y}$ and mirror ${\cal M}_x$, ${\cal M}_y$ symmetries are broken due to the incompatible transformation of the two superconducting orders under the symmetries. The symmetry breaking splits each node into a pair of Weyl nodes with opposite charge $\pm 1$. The Weyl nodes have the same $k_x$ momentum and are located in the $k_y = 0$ plane. Furthermore, due to TRSB, the Majorana Dirac cones on mirror ${\cal M}_z$ symmetry-breaking surfaces [such as the $(0,-1,1)$ surface] acquire a mass term opening a spectral gap on the surfaces [see Fig.~\ref{fig:sketch_topo}(c)]. Surfaces related by mirror ${\cal M}_z$ symmetry have an opposite sign of the mass term. At the same time, the zero-energy flat bands acquire a dispersion turning it into a chiral Majorana mode \cite{ReadGreen00}. The chiral Majorana mode is protected by the massive Dirac theories with opposite sign on mirror ${\cal M}_z$ symmetric hinges. This system is a second-order topological nodal superconductor of type (iii) \cite{SimonPRB2022}. In addition to the massive Dirac cones around the center of the BZ, the surface hosts Fermi arcs of Bogoliubov quasiparticles connecting to the projection of the bulk Weyl nodes around $(\pm k_F, 0, 0)^{\rm T}$. As the total charge of Weyl nodes in each half-space with positive or negative $k_x$ is zero, the Fermi arcs connect only Weyl nodes within each half space. Similarly to the case with $\mathrm{B_{3u}}$ superconducting order, step edges on mirror ${\cal M}_z$ symmetry-breaking surfaces [such as the $(0,-1,1)$ surface] host chiral Majorana modes if the mass terms on adjacent surfaces have opposite sign [Fig.~\ref{fig:sketch_topo}(d)]. 

Ref.~\cite{ShishidouEA21} discusses $\mathrm{B_{3u}} + i \mathrm{B_{2u}}$ pairing as the most likely candidate for TRSB multi-component order in ${\rm UTe}_2$. This work highlights the appearance of Weyl nodes, but does not discuss the second-order topology described here. 

{\em ${\rm B}_{2\mathrm{u}} + i \varepsilon \mathrm{B_{1u}}$ superconducting order and other permutations.---} The discussion for superconducting orders with other combinations of $\mathrm{B}_{k\mathrm{u}},\ k = 1,2,3$ symmetry can be obtained by a permutation of coordinates. Here, we explicitly discuss the experimentally relevant case of ${\rm B}_{2\mathrm{u}} + i \varepsilon \mathrm{B_{1u}}$ symmetry~\cite{SLee2023,Theuss2023}.

With dominant ${\rm B}_{2\mathrm{u}}$ symmetric order, the nodes are located around $(0, \pm k_F, 0)^{\rm T}$. An admixture with a ${\rm B}_{1\mathrm{u}}$ symmetric order and relative phase breaks time-reversal, rotation ${\cal R}_y$, ${\cal R}_z$, and mirror ${\cal M}_y$, ${\cal M}_z$ symmetries. As the $(0,-1,1)$ surface preserves the mirror ${\cal M}_x$ symmetry, it hosts a Majorana Dirac surface cone where the crossing at $k_x = 0$ is protected by mirror ${\cal M}_x$ symmetry. It coexists with the projection of the bulk nodes onto the surface BZ. A similar result holds for ${\rm B}_{1\mathrm{u}} + i \varepsilon \mathrm{B_{2u}}$. 

The other permutations ${\rm B}_{3\mathrm{u}} + i \varepsilon \mathrm{B_{1u}}$, ${\rm B}_{3\mathrm{u}} + i \varepsilon \mathrm{B_{2u}}$, and ${\rm B}_{2\mathrm{u}} + i \varepsilon \mathrm{B_{3u}}$ break mirror ${\cal M}_x$ symmetry. Therefore, $(0,-1,1)$ surfaces with these orders are gapped except for the projections of the bulk Weyl nodes and corresponding Fermi arcs. 

{\em ${\rm B}_{3\mathrm{u}} + i \varepsilon {\rm A}_{\mathrm{u}}$ superconducting order.---} Including a small admixture of superconducting order with ${\rm A}_{\mathrm{u}}$ symmetry with relative phase breaks time-reversal, rotation ${\cal R}_{y}$, ${\cal R}_{z}$, and mirror ${\cal M}_y$, ${\cal M}_z$ symmetries. With these broken symmetries, the slice at $0 < |k_x| < \pi$ becomes topologically trivial, such that the nodes at $(\pm k_F, 0, 0)^{\rm T}$ are no longer topologically protected. Instead, the admixture splits each node into four Weyl nodes with cancelling total charge away from $(\pm k_F, 0, 0)^{\rm T}$ as shown in Fig.~\ref{fig:Gaps}. Each Weyl node has charge $\pm 1$. The slice at $k_x = 0$ remains topologically non-trivial characterized by a mirror Chern number equal to one. The mirror Chern number indicates the presence of mirror-symmetry protected Majorana Dirac cones on mirror ${\cal M}_x$ symmetric surfaces. These Majorana Dirac cones are realized on the $(0, -1, 1)$ crystal surface. If the bulk is fully gapped or the nodes do not lie within the $k_y = 0$ or $k_z = 0$ plane, then mirror ${\cal M}_x$-symmetric hinges also support chiral Majorana modes. These are associated to the mirror Chern number.

{\em ${\rm B}_{1\mathrm{u}} + i \varepsilon {\rm A}_{\mathrm{u}}$ and ${\rm B}_{2\mathrm{u}} + i \varepsilon {\rm A}_{\mathrm{u}}$ superconducting order.---} For ${\rm B}_{1\mathrm{u}} + i \varepsilon {\rm A}_{\mathrm{u}}$ [${\rm B}_{2\mathrm{u}} + i \varepsilon {\rm A}_{\mathrm{u}}$] mixed superconducting order, the nodes are along the $k_z$ [$k_y$] direction and the remaining mirror symmetry is ${\cal M}_z$ [${\cal M}_y$]. In these cases, the $(0, -1, 1)$ crystal surface is gapped except for the Fermi arcs around the projection of the Weyl nodes onto the surface BZ. 

{\em ${\rm B}_{3\mathrm{u}} + i {\rm A}_{\mathrm{u}}$ superconducting order with $\vec{d}(\bo{k}) = (0, c_1 k_z + i c_2 k_y, 0)^{\rm T}$.---} We furthermore consider a TRSB order parameter of the specific form $\vec{d}(\bo{k}) = (0, c_1 k_z + i c_2 k_y, 0)^{\rm T}$. This order parameter describes a Weyl superconductor \cite{MengPhysRevB2012Aug} with Weyl nodes with charge $\pm 2$ at $\bo{k} = (\pm k_F, 0, 0)^{\rm T}$. The projections of the Weyl nodes with positive and negative $k_x$ onto the surface BZ are connected by Fermi arcs of Bogoliubov quasiparticles. This connectivity distinguishes them from Weyl nodes discussed above for the other nodal phases. All eigenstates are two-fold degenerate due to a ${\rm SU}(2)$ spin-rotation symmetry. The Bogoliubov Fermi arcs are realized on the $(0, -1, 1)^{\rm T}$ crystal surface and connect to the Weyl nodes at $\bo{k} = (\pm k_F, 0, 0)^{\rm T}$. For combinations ${\rm B}_{1\mathrm{u}} + i {\rm A}_{\mathrm{u}}$ and ${\rm B}_{2\mathrm{u}} + i {\rm A}_{\mathrm{u}}$, Weyl nodes can be obtained along the $k_z$ and $k_y$ direction, respectively.

\subsubsection{Gapped phases}
\label{sec:TopPhases_gapped}
{\em ${\rm A}_{\rm u}$ superconducting order.---} A spherical Fermi surface with ${\rm A}_{\rm u}$ superconducting order parameter $\vec{d}(\bo{k}) = (c_1 k_x, c_2 k_y, c_3 k_z)$ is fully gapped and realizes a strong topological superconductor in class DIII hosting surface Majorana Dirac cones. 

{\em ${\rm A}_{\rm u} + i \varepsilon {\rm B}_{3 {\rm u}}$ superconducting order.---} A small ($|\varepsilon| \ll 1$) admixture of superconducting order with ${\rm B}_{3 {\rm u}}$ symmetry and relative phase to an ${\rm A}_{\rm u}$ superconducting order breaks time-reversal symmetry as well as mirror ${\cal M}_y$ and ${\cal M}_z$ and rotation ${\cal R}_{y}$ and ${\cal R}_{z}$ symmetries. This gaps the Majorana Dirac surfaces cones and turns the system into a second-order topological superconductor hosting chiral Majorana modes on mirror ${\cal M}_x$ symmetric hinges. Similar to the nodal phase with $\mathrm{B_{3u}} + i \varepsilon {\rm A}_{\rm u}$ superconducting order, step edges on a mirror-symmetry breaking surface host a chiral Majorana mode if the microscopic surface theories on the two sides of the hinge lead to an opposite sign of the mass term of the massive surface Dirac Hamiltonian. The $(0, -1, 1)$ surface preserves the ${\cal M}_x$ mirror symmetry and thereby hosts a Majorana Dirac surface cone protected by mirror ${\cal M}_x$ symmetry.

{\em ${\rm A}_{\rm u} + i \varepsilon {\rm B}_{1 {\rm u}}$ and ${\rm A}_{\rm u} + i \varepsilon {\rm B}_{2 {\rm u}}$ superconducting order.---} For a ${\rm B}_{1 {\rm u}}$ [${\rm B}_{2 {\rm u}}$] symmetric admixture with relative phase, a mirror ${\cal M}_z$ [${\cal M}_y$] symmetry remains. Then, the system becomes a second-order topological superconductor with chiral Majorana modes on mirror ${\cal M}_z$ [${\cal M}_y$] symmetric hinges. In both cases, the $(0, -1, 1)$ surface is gapped. 

\subsubsection{Symmetry-based indicator for DFT+$U$ bandstructure}
\label{sec:UTe2_topology_SI}
The topology of a BdG Hamiltonian resulting from introducing small superconducting pairing in a metal with given bandstructure can be analyzed using the weak-pairing expressions of symmetry-based indicators~\cite{OnoPhysRevRes2019Aug,SkurativskaPhysRevRes2020Jan,GeierPhysRevB2020Jun,OnoPhysRevRes2021May}. This analysis requires that the pairing strength is small compared to the relevant energy scales of the normal-state Hamiltonian such that no gaps of the resulting BdG Hamiltonian are closed as the pairing strength is increased. It is further assumed that the BdG Hamiltonian is gapped at the high-symmetry momenta. The symmetry-based indicators can detect fully-gapped topological phases, as well as topological nodal phases \cite{TangPhysRevLett2022Jul}.

Here, we apply the theory of symmetry-based indicators to study the topological phases resulting from the normal state bandstructure of ${\rm UTe}_2$ obtained from our DFT+$U$ study as summarized in Sec.~\ref{sec:electronicUTe2}. We calculate the symmetry-based indicator $z_3$ of strong topological phases in 3D, inversion-symmetric, odd-parity superconductors in class DIII using its weak-pairing expression~\cite{GeierPhysRevB2020Jun}
\begin{equation}
    z_3 = \sum_{\textbf{k}_{\rm s}} \left( n^{\textbf{k}_{\rm s}}_+ |_{\rm occ.} - n^{\textbf{k}_{\rm s}}_- |_{\rm occ.} \right) (-1)^{2 (h + k + l)} \mod 8
    \label{eq:z_3_WeakPairing}
\end{equation}
where the sum runs over all eight inversion-symmetric momenta $\textbf{k}_{\rm s} = h \textbf{G}_1 + k \textbf{G}_2 + l \textbf{G}_3$ written in terms of the primitive reciprocal lattice vectors $\textbf{G}_j,\ j = 1,2,3$, and $n^{\textbf{k}_{\rm s}}_\alpha |_{\rm occ.}$ is the number of occupied Kramers pairs with inversion parity $\alpha$ at $\textbf{k}_{\rm s}$ of the normal state bandstructure~\footnote{In Eq.~\eqref{eq:z_3_WeakPairing} we further used that terms $n^{\textbf{k}_{\rm s}}_-$ counting the total number of bands with odd inversion parity can be chosen to be an integer multiple of eight by an appropriate energy cut-off such that $n^{\textbf{k}_{\rm s}}_- \mod 8 = 0$ at all inversion-symmetric momenta $\textbf{k}_{\rm s}$.}.

The symmetry-based indicator $z_3$ counts the maximum number of Majorana Dirac surface cones that are present on any surface modulo eight~\cite{GeierPhysRevB2018May}. With inversion symmetry, there are both first-order topological phases hosting a $\mathbb{Z}$ number of Majorana Dirac surface cones, as well as second- and third-order topological phases as a result of Majorana Dirac surface cones gapping out in sets of two and four to produce a helical Majorana hinge mode and a Kramers pair of Majorana corner states, respectively. This leads to an ambiguity in the topological phases indicated by $z_3$, such that $z_3 = 2$ can describe a second-order or a first-order topological superconducting phase hosting two Majorana Dirac surface cones, $z_3 = 3$ can describe a second-order mixed with a first-order topological superconducting phase hosting a single Majorana Dirac surface cone or a first-order topological superconductor hosting three Majorana Dirac surface cones, and so on. For $z_3$ odd, the surfaces are always gapless due to the presence of an odd number of Majorana Dirac surface cones. 

In point group $D_{2\mathrm{h}}$, the crystalline symmetries in addition to inversion symmetry impact the topological classification and thereby the interpretation of the symmetry-based indicator $z_3$. With $\mathrm{A_u}$ symmetric pairing, the topological classification contains first- and higher-order topological phases separately~\cite{ShiozakiProgTheorExpPhys2022Apr}, as in point group $C_i$. In this case, the interpretation of $z_3$ is the same as in $C_i$ as summarized above. With $\mathrm{B_{1u}}$, $\mathrm{B_{2u}}$, or $\mathrm{B_{3u}}$ symmetric pairing, the first-order topological phase is forbidden by the rotation and mirror symmetries \cite{ShiozakiProgTheorExpPhys2022Apr}. This is reflected by our analysis in Sec.~\ref{sec:TopPhases_nodal} where we found that single Majorana Dirac surfaces cones occur together with a bulk node protected by a two-dimensional second-order topological invariant. As a consequence, for $\mathrm{B_{1u}}$, $\mathrm{B_{2u}}$, or $\mathrm{B_{3u}}$ symmetric pairing, odd values of $z_3$ indicate a second-order topological nodal superconductor as described in Sec.~\ref{sec:TopPhases_nodal}. For these phases, including an additional, infinitesimal superconducting order breaking time-reversal symmetry leads to the phenomenology described in Sec.~\ref{sec:TopPhases_nodal} and \ref{sec:TopPhases_gapped}. 

\begin{figure}
    \centering
    \includegraphics[width=\columnwidth]{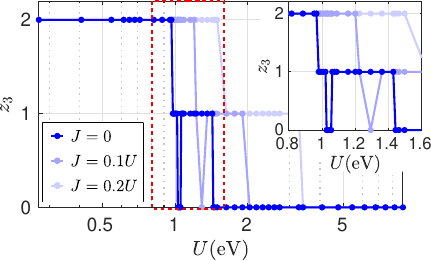}
    \caption{Symmetry-based indicator $z_3$ calculated from Eq.~\eqref{eq:z_3_WeakPairing} and the DFT+$U$ bandstructure for ${\rm UTe}_2$ as a function of Hubbard repulsion $U$ and for Hund's exchange interaction $J = 0$, $0.1\, U$, and $0.2\, U$, as discussed in Sec.~\ref{sec:electronicUTe2}. The inset shows a blowup of the data at an intermediate $U$ range as indicated by the red dashed box in the main panel.}
    \label{fig:z3_DFT+$U$}
\end{figure}

Figure~\ref{fig:z3_DFT+$U$} shows the symmetry-based indicator $z_3$ from Eq.~\eqref{eq:z_3_WeakPairing} obtained for the DFT+$U$ bandstructure calculations summarized in Sec.~\ref{sec:electronicUTe2} as a function of Hubbard repulsion $U$. For the experimentally relevant range of moderate $U \approx 0.97$-$1.44~{\rm eV}$ \cite{IshizukaEA19, ShishidouEA21}, we find $z_3 = 1$, except for a small region around $U \approx 1.05~{\rm eV}$ where an additional Fermi pocket around $\Gamma$ appears [see Sec.~\ref{sec:electronicUTe2}]. This result is consistent with the shape of the Fermi surface obtained from the DFT+$U$ calculations for moderate $U$: the electron-like Fermi surface enclosing the $X$ point [blue and yellow surface in Fig.~\ref{fig:bands_FS}] can be deformed into a spherical Fermi surface around the $X$ point and a cylindrical Fermi surface enclosing the other time-reversal symmetric momenta at the Brillouin zone boundary without crossing any time-reversal symmetric momenta but allowing cutting and gluing of surfaces away from these high-symmetry momenta. This set of rules ensures that the $z_3$ symmetry-based indicator dictating the ground state topology after introducing superconducting pairing remains invariant under the deformations~\footnote{Notice that this set of allowed deformations is distinct from the continuous deformations considered when studying homotopy equivalence of shapes. In addition to continuous deformations, we allow to cut and glue surfaces away from the time-reversal symmetric momenta. At the same time, we impose the additional rule that no Fermi surface may cross the time-reversal symmetric momenta.}. For the spherical Fermi pocket, our analysis in Sec.~\ref{sec:TopPhases_nodal} and \ref{sec:TopPhases_gapped} applies. The remaining cylindrical Fermi surfaces are irrelevant for the symmetry-based indicator $z_3$ as they enclose pairs of inversion-symmetric momenta with cancelling contributions. The region for small $U \lesssim 0.97~{\rm e V}$ where $z_3 = 2$ is experimentally not relevant because the DFT+$U$ calculations yield an insulator in this range.

%
\section{Discussion and Conclusions}
\label{sec:Conclusion}
%
Through a general Ginzburg--Landau analysis, we have provided a discussion of two-component spin-triplet superconducting orders and their associated allowed composite structure. This includes criteria for when TRSB non-unitary order emerges at the second transition. The associated secondary specific heat transition may exhibit the peculiar feature that it is vanishingly small or even drops upon entering a TRSB non-unitary state. These are general properties under such circumstances, but we have exemplified these results through superconducting states relevant for UTe$_2$, a heavy-fermion compound under considerable current interest due to its strong evidence for topological spin-triplet superconductivity. 

Some experiments on UTe$_2$ are most consistent with TRSB non-unitary two-component triplet superconductivity~\cite{Jiao2020,HayesEA21,Bae2021,KotaEA23}. In that context, our results provide a possible resolution to the lacking observation of a second specific heat jump. Other experiments are more consistent with a condensate consisting of a single component of the 1D triplet irreps, and indeed the simplest explanation of the specific heat behavior of high-quality UTe$_2$ samples is that the order is single-component and belongs to one of the odd-parity irreps possessing point nodes.

This current puzzle of the detailed pairing structure of UTe$_2$ motivated us to pursue also the topological properties of the different superconducting states under consideration for this material. For a single-band spherical Fermi surface, single-component nodal phases with $\mathrm{B_{1u}}$, $\mathrm{B_{2u}}$, or $\mathrm{B_{3u}}$ pairing host surface Majorana Dirac cones and flat zero-energy bands at hinges in addition to the bulk nodes. Within our scope of analyzing the topological properties of TRSB two-component orders under the assumption that the pairing strength of one order is much weaker than the other, the observed gapped $(0,-1,1)$ surface with chiral modes at step edges~\cite{Jiao2020} are consistent with $\mathrm{B_{3u}} + i \varepsilon \mathrm{B_{2u}}$, $\mathrm{B_{2u}} + i \varepsilon \mathrm{B_{3u}}$, $\mathrm{B_{3u}} + i \varepsilon \mathrm{B_{1u}}$, $\mathrm{B_{1u}} + i \varepsilon \mathrm{B_{3u}}$, $\mathrm{B_{1u}} + i \varepsilon \mathrm{A_{u}}$, and $\mathrm{B_{2u}} + i \varepsilon \mathrm{A_{u}}$ pairing. For these orders, the node of the dominant $\mathrm{B}_{j\mathrm{u}},\ j = 1,2,3, $ order splits into multiple Weyl nodes with cancelling charge. The $(0,-1,1)$ surface is gapped, except for Fermi arcs at large momentum around the projection of the Weyl nodes. Step edges between chemically distinct surfaces may host chiral modes. Alternatively, an increased density of states at step edges could be related to Fermi arcs stemming from the Weyl nodes. A fully-gapped spin-triplet superconducting state points toward dominant $\mathrm{A_u}$ pairing, with potential $\mathrm{B_{1u}}$ or $\mathrm{B_{2u}}$-symmetric admixture breaking TRS. Single-component $\mathrm{A_u}$ pairing host surface Majorana Dirac cones that would gap out under a $\mathrm{B_{1u}}$ or $\mathrm{B_{2u}}$ TRSB admixture. The remaining orders host Majorana Dirac cones on the $(0,-1,1)$ surface protected by mirror symmetry also when TRS is broken. Our model of a spherical Fermi surface is motivated by DFT+$U$ band structure calculations from which we calculate a symmetry-based indicator supporting the applicability of our results.

The topological phases discussed for the various odd-parity superconducting orders may further bind anomalous modes to vortices. It is well-known that the first-order topological superconductor as we found for $\mathrm{A_u}$ pairing hosts helical Majorana modes at vortices~\cite{ChiuEA16}. For nodal $\mathrm{B}_{j\mathrm{u}}$ pairing, we also expect helical Majorana modes at vortex lines that are not perpendicular to the axis connecting the bulk nodes. For the fully-gapped second-order topological superconducting state with TRSB $\mathrm{A_u} + i \varepsilon \mathrm{B}_{j\mathrm{u}}$ pairing we expect that vortex line ends may bind Majorana zero modes because the gapped surfaces are described by massive Dirac theories in Cartan class D, however we expect their presence to depend on microscopic details. Similarly, the second-order topological phases in the remaining nodal superconducting orders may host Majorana zero modes at the ends of vortex lines. For nodal phases, Majorana zero modes are not topologically protected because of hybridization with the gapless bulk or Fermi arc surface states. Besides vortices, also topological lattice defects, such as dislocations, disclinations, and grain boundaries, host anomalous modes directly related to the crystalline bulk topology and superconducting order~\cite{TeoPhysRevB2010Sep, ThorngrenPhysRevX2018Mar, GeierSciPostPhys2021Apr}. The presence of absence of such anomalous defect modes, and their signatures in  surface-probe measurements~\cite{WangScience2018Oct,KongNatPhys2019Nov,ZhuScience2020Jan,SbierskiPhysRevB2022Jul, NayakNatPhys2021Dec, BatabyalSciAdv2016Aug, HowardNatCommun2021Jul, NayakSciAdv2019Nov, QueirozPhysRevLett2019Dec}, provide an interesting future research direction that may help further pin down the bulk topology and pairing symmetry of the fascinating heavy-fermion material UTe$_2$.

\begin{acknowledgments}
We acknowledge useful discussions with Morten H.~Christensen, Hans Christiansen, J.~C.~S{\'e}amus Davis, and Takasada Shibauchi.
H.S.R.~was supported by research Grant No.~40509 from VILLUM FONDEN. A.K.~acknowledges support by the Danish National Committee for Research Infrastructure (NUFI) through the ESS-Lighthouse Q-MAT. M.G.~acknowledges support by the European Research Council (ERC) under the European Union’s Horizon 2020 research and innovation program under grant agreement No.~856526, and from the Danish National Research Foundation, the Danish Council for Independent Research $\vert$ Natural Sciences, and the German Research Foundation under the Walter Benjamin program (Grant Agreement No.~526129603).
\end{acknowledgments}

\appendix

%
\section{Intermezzo: two-component orders and microscopic evaluation}
\label{app:Intermezzo}
\begin{figure}[tb!h]
	\centering
	\includegraphics[width=0.55\linewidth]{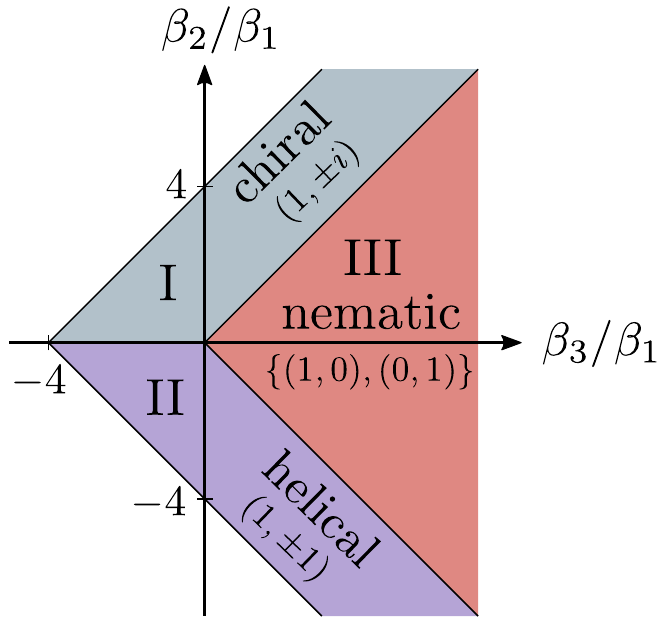}
    \caption{Phase diagram of a two-component order parameter belonging to, e.g., $\mathrm{E_u}$ of $D_{4\mathrm{h}}$. The chiral phase breaks time-reversal symmetry, and the nematic and helical phases break rotational symmetry, $C_4 \to C_2$.}
	\label{fig:TwoComponent}
\end{figure}

We consider a spin triplet superconductor with an order parameter, $\vec{\eta} \in \mathbb{C}^2$, belonging to the two-dimensional irreducible representation $\mathrm{E_u}$ of the tetragonal point group $D_{4\mathrm{h}}$. The symmetry-consistent Ginzburg--Landau theory in this case takes the form~\cite{SigristUeda91}
\begin{align}
    \pazocal{F}_{\mathrm{E_u}}[\vec{\eta}] &= \alpha(T) \lvert \vec{\eta} \rvert^2 + \beta_1 \lvert \vec{\eta} \rvert^4 + \frac12 \beta_2\left( (\eta_x^{\ast})^2 \eta_y^2 + \eta_x^2 (\eta_y^{\ast})^2 \right) \notag\\
    &\hspace{40pt} + \beta_3 \lvert \eta_x \rvert^2 \lvert \eta_y \rvert^2.
    \label{eq:FreeEnergyTwoComponent}
\end{align}
The phase diagram of this model is easily derived from the parametrization $\vec{\eta} = \eta_0(\cos{\phi},~\exp(i\theta)\sin{\phi})^{\mathrm{T}}$ which upon insertion in Eq.~\eqref{eq:FreeEnergyTwoComponent} gives the following criteria for thermodynamic stability (requiring the free energy to be lower bounded):
\begin{equation}
\begin{aligned}
    \beta_1 &> 0, \\
    4\beta_1 + \beta_3 \pm \beta_2 &> 0.
    \label{eq:ThermodynamicStability}
\end{aligned}
\end{equation}
Minimizing Eq.~\eqref{eq:FreeEnergyTwoComponent} for parameters satisfying these criteria result in the three standard phases shown in Fig.~\ref{fig:TwoComponent}. A priori, all three phases appear realizable. 

We contrast the above theory to a microscopic approach in which the free energy coefficients are evaluated from the diagrammatic loop expansion, following Gor'kov~\cite{Gorkov59} and explained in detail elsewhere~\cite{ScheurerEA20, WagnerEA21}. In the absence of spin-orbit coupling we can without loss of generality take the order parameter to be $\vec{d} = \hat{z}(\eta_x f_x + \eta_y f_y)$, where $f_x$ and $f_y$ are momentum dependent form factors normalized by their maximal absolute value. The resulting microscopic free energy is:
\begin{equation}
\begin{aligned}
    \pazocal{F}_{\mathrm{E_u}}[\vec{\eta}] &= \tilde{\alpha}(T,T_c) \left( \lvert \eta_x \vert^2 + \lvert \eta_y \rvert^2 \right) + \tilde{\beta}_1\left( \lvert \eta_x \rvert^4 + \lvert \eta_y \rvert^4 \right) \\
    &\hspace{20pt} + \tilde{\beta}_2 \left[ 4\lvert \eta_x \rvert^2 \lvert \eta_y \rvert^2 + (\eta_x^{\ast})^2\eta_y^2 + \eta_x^2(\eta_y^{\ast})^2 \right], \\
    \tilde\alpha(T,T_{c}) &= - V  \int \f{\D^d \bo{p}}{(2\pi)^d}~\Big( \f{\tanh\left[ \xi(\bo{p}) / (2T) \right] }{2\xi (\bo{p})} \\
    &\hspace{40pt} -\f{\tanh\left[ \xi(\bo{p}) / (2T_{c}) \right]}{2\xi(\bo{p})} \Big) f^2_{j}(\bo{p}), \\
    \begin{bmatrix}
        \tilde{\beta}_1 \\
        \tilde{\beta}_2
    \end{bmatrix} &= \f{V}{2T^3} \int \f{\D^d \bo{p}}{(2\pi)^d}~h( \xi(\bo{p})/ T ) \begin{bmatrix}
    f_j^4(\bo{p}) \\
    f_x^2(\bo{p})f_y^2(\bo{p})
    \end{bmatrix},
\end{aligned}
\end{equation}
where $\xi(\bo{p})$ is the normal-state dispersion, $V$ is the unit cell volume, where $j = x, y$ are equal by symmetry, and finally where the function $h$ is given by
\begin{equation}
h(x) \equiv \frac{\sinh{x}-x}{4x^3(1+\cosh{x})}.
\label{eq:hfunc}
\end{equation}

The Cauchy--Schwarz inequality tells us that $\tilde{\beta}_1^2 \geqslant \tilde{\beta}_2^2 \Rightarrow \tilde{\beta}_1 \geqslant \tilde{\beta}_2$. Comparing with the coefficients of the phenomenological theory in Eq.~\eqref{eq:FreeEnergyTwoComponent}, we see that $\beta_1 = \tilde{\beta}_1 \geqslant 0$, $\beta_2 = 2\tilde{\beta}_2 \geqslant 0$, and $\beta_3 = 4\tilde{\beta}_2 - 2\tilde{\beta}_1$. From the above expressions combined with the Cauchy--Schwarz inequality we find that $\beta_2 = 2\tilde{\beta}_2 \geqslant 0$ and $\beta_3 = 2\tilde{\beta}_2 +2(\tilde{\beta}_2 - \tilde{\beta}_1) \leqslant 2\tilde{\beta}_2 = \beta_2$. Placing this in the phase diagram of Fig.~\ref{fig:TwoComponent}, we reach the following conclusion for the microscopic theory: \\ 

\noindent \textbf{Theorem:} \emph{At the mean-field level of a single-band superconductor with crystal point group $D_{4\mathrm{h}}$ and an order parameter belonging to the irreducible representation $\mathrm{E_u}$, the chiral TRSB phase (I) of Fig.~\ref{fig:TwoComponent}, i.e.,~$\vec{d} = \hat{z} \Delta_0 (f_x + i f_y)$ is favoured when using the loop expansion to evaluate the Ginzburg--Landau coefficients.} \\

This simple fact does not appear to be commonly pointed out in the literature, although Ref.~\cite{SaulsUPt3} analogously mentions that the weak-coupling limit with $D_{6\mathrm{h}}$ symmetry permits a constant ratio between the only two quartic-order coefficients of the theory. It is an interesting question to explore the conditions for this result to break down. 

%
\section{Details of thermodynamic spin-triplet double transitions}
\label{app:Entropy}
%
%
\subsection{Entropy of the double transitions}
\begin{figure*}[tb]
	\centering
	\includegraphics[width=0.8\linewidth]{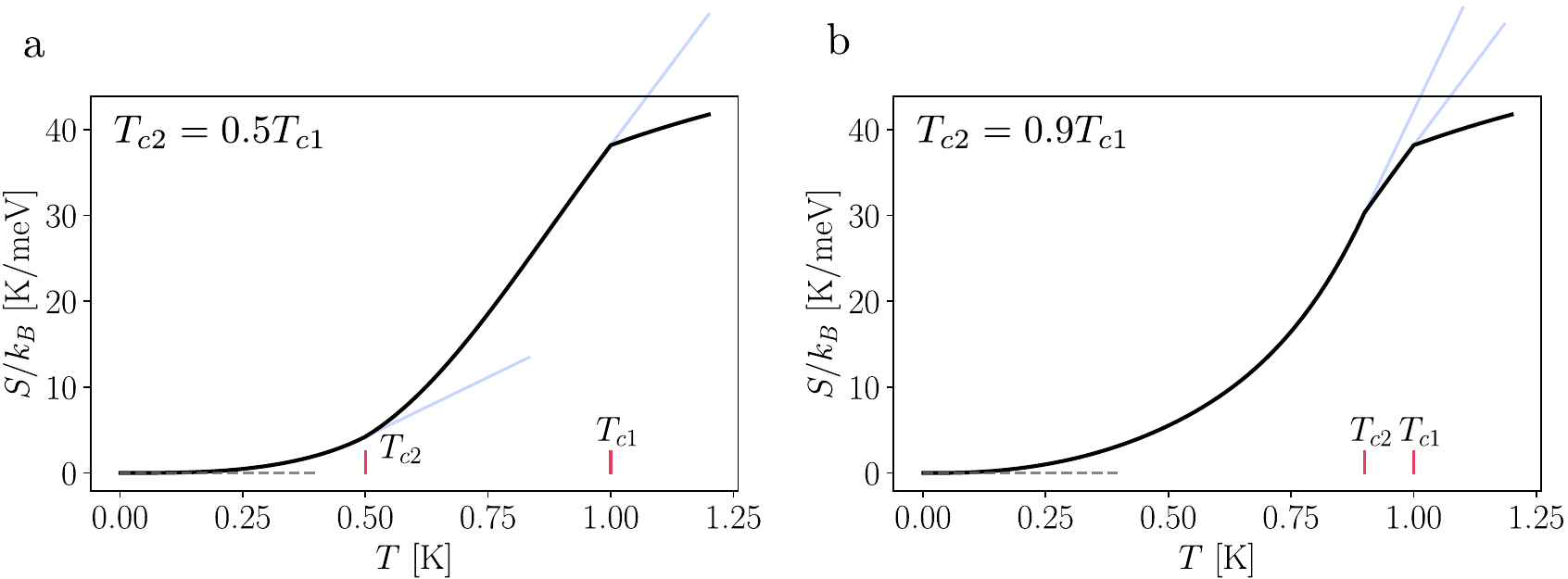}
    \caption{The entropy calculated from Eq.~\eqref{eq:Entropy} for a spherical Fermi surface with $k_F^2/(2m) = 1~$meV for $T_{c1} = 1.0~$K and (a) $T_{c2} = 0.5~$K, and (b) $T_{c2} = 0.9~$K. The transparent blue lines show the slope of $S$ right below the transition temperatures at which the entropy is non-analytic. In panel (a) $\mathrm{sign}[\frac{\partial S}{\partial T}\big\rvert_{T_0^{-}} - \frac{\partial S}{\partial T}\big\rvert_{T_{0}^+}]$ is positive at $T_0 = T_{c1}$ and negative at $T_0 = T_{c2}$. In panel (b) this quantity is positive at both $T_0 = T_{c1}$ and $T_0 = T_{c2}$. }
	\label{fig:Entropy}
\end{figure*}

To elaborate on how the heat capacity anomalies observed in Fig.~\ref{fig:HeatCapacity} manifest in the entropy, we have calculated the entropy from Eq.~\eqref{eq:Entropy} for $T_{c2} = 0.5 T_{c1}$ and $T_{c2} = 0.9 T_{c1}$ in Fig.~\ref{fig:Entropy}. For these critical temperatures the calculations in Fig.~\ref{fig:HeatCapacity} resulted in negative and positive heat capacity anomalies, respectively. 

As stated in the main text, the non-analyticity of $S$, i.e., $\mathrm{sign}[\frac{\partial S}{\partial T}\big\rvert_{T_{c2}^-} - \frac{\partial S}{\partial T}\big\rvert_{T_{c2}^+}]$ dictates whether the second heat capacity discontinuity is positive or negative. This is consistently confirmed when calculating the entropy directly from Eq.~\eqref{eq:Entropy}, as shown in Fig.~\ref{fig:Entropy}. With $T_{c2} = 0.5 T_{c1}$ we find that $\frac{\partial S}{\partial T}\big\rvert_{T_{c2}^-} - \frac{\partial S}{\partial T}\big\rvert_{T_{c2}^+} < 0$ (consistent with $\Delta C(T_{c2}) < 0$), whereas for $T_{c2} = 0.9 T_{c1}$ we find $\frac{\partial S}{\partial T}\big\rvert_{T_{c2}^-} - \frac{\partial S}{\partial T}\big\rvert_{T_{c2}^+} > 0$ (consistent with $\Delta C(T_{c2}) > 0$).

Returning to the result of Eq.~\eqref{eq:Jump} we next normalize by the primary jump at $T_{c1}$ to obtain
\begin{widetext}
\begin{equation}
\frac{\Delta C(T_{c2})}{\Delta C(T_{c1})} \frac{T_{c1}}{T_{c2}} = \left( \frac{T_{c1}}{T_{c2}} \right)^2 \varepsilon^2 \frac{ \frac{1}{4k_B T_{c2}} \int_{-\infty}^{\infty} \D \xi \big\langle \sech^{2}\big( \frac{\sqrt{\xi^2+g^2} }{2k_B T_{c2}} \big) \big[  \lvert \Vec{d}_2 \rvert^2 - \frac{\Delta_0^2}{k_B T_{c2}} \frac{\lvert \vec{d}_1 \times \vec{d}_2 \rvert^2 }{\sqrt{\xi^2 + g^2}} \left( 1-\frac{T_{c2}}{T_{c1}} \right) \tanh \big( \frac{\sqrt{\xi^2+g^2} }{2k_B T_{c2}} \big) \big] \big\rangle_{\mathrm{FS}} }{\langle \lvert \Vec{d}_1 \rvert^2 \rangle_{\mathrm{FS}}},
\label{eq:JumpRatio}
\end{equation} 
\end{widetext}
where still $g = \Delta_0 \lvert \Vec{d}_1 \rvert \sqrt{1-T_{c2}/T_{c1}}$. For infinitesimal splittings, $T_{c2} \to T_{c1}$, this result simplifies considerably: $\Delta C(T_{c2} = T_{c1})/\Delta C(T_{c1}) = \varepsilon^2 \langle \lvert \Vec{d}_2 \rvert^2 \rangle_{\mathrm{FS}} / \langle \lvert \Vec{d}_1 \rvert^2 \rangle_{\mathrm{FS}}$, i.e., the order parameter anisotropy ratio. This explains the results of Fig.~\ref{fig:HeatCapacity} at $T_{c2} = 1$~K: the dashed lines have  $\langle \lvert \vec{d}_1 \rvert^2 \rangle_{\mathrm{FS}} = \langle \lvert \vec{d}_2 \rvert^2 \rangle_{\mathrm{FS}}$, whereas the full line clearly has $\langle \lvert \vec{d}_1 \rvert^2 \rangle_{\mathrm{FS}} < \langle \lvert \vec{d}_2 \rvert^2 \rangle_{\mathrm{FS}}$ at this temperature.

\subsection{Critical exponents beyond mean field}
So far we have considered the case of mean-field critical exponents ($\beta = \frac12$) in the order parameter temperature dependence. The numerical value of the corresponding critical exponent in the XY universality class in 3D is $\beta_{\mathrm{XY}} \approx 0.3485(2)$~\cite{CampostriniEA01}. Let us therefore focus on the derivation of the unusual negative term in Eq.~\eqref{eq:Jump} from the more general order parameter ansatz of
\begin{equation}
\begin{aligned}
    \vec{d}(T, \bo{k}) &= \Delta_{0} [ ( 1- T/T_{c1} )^a \vec{d}_{1}(\bo{k}) \\
    &\hspace{40pt} + i \varepsilon ( 1 - T/T_{c2} )^b \vec{d}_{2}(\bo{k}) ],
\label{eq:GeneralAnsatz}
\end{aligned}
\end{equation}
where $0 < a,~b < 1$. Still expressing $T_{c2}^- = T_{c2} - \delta T$, we have $\partial \lvert \vec{d}^{\ast} \times \vec{d} \rvert /( \partial T) \big\rvert_{T_{c2}^-} \sim (\delta T)^{b-1}$. The other factor appearing in Eq.~\eqref{eq:HeatCapacityMassage}, on the other hand, behaves as $\sum_{\sigma} \sigma\sech^2(E_{\sigma}/(2k_B T_{c2}^-)) \sim (\delta T)^b$, so the general form of the heat capacity integrand is
\begin{equation}
    \delta c = A \lvert \vec{d}_2 \rvert^2 - B \lvert \vec{d}_1 \cross \vec{d}_2 \rvert^2 (\delta T)^{2b-1},
    \label{eq:deltacGeneral}
\end{equation}
where $A$ and $B$ are numerical factors. Hence, the mean-field case of $b = \frac12$ is peculiar in the sense that it separates a vanishing result ($b > \frac12$) from a formally divergent result ($b < \frac12$). In principle, non-unitary secondary transitions in the entire range of $0 < b \leq \frac12$ can accommodate the anomalous heat capacity signature.

%
\section{Single-component non-unitary transition}
\label{app:SingleNonUnitary}
%
Consider the case of a single-component non-unitary order parameter onsetting at $T_c$:
\begin{equation}
    \vec{d}(T,\bo{k}) = \Delta_0 \sqrt{1-T/T_c} \; \hat{d}(\bo{k}).
    \label{eq:SingleNonUnitary}
\end{equation}
The associated (squared) gaps are given by $\lvert \Delta_{\sigma} \rvert^2 = \Delta_0^2 \left( 1-\frac{T}{T_c} \right) \left( \lvert \hat{d}\rvert^2 + \sigma \lvert \hat{d}^{\ast} \times \hat{d} \rvert \right)$. Employing this to calculate the specific heat jump $\Delta C(T_c)$, using Eq.~\eqref{eq:Specificheat}, now straightforwardly results in 
\begin{equation}
\begin{aligned}
    \Delta C(T_c) &\equiv C(T_c^{-}) - C(T_c^+) \\
 & \! \!\!\!\!\! \!\!= \left( \gamma_n T_c + \frac{\Delta_0^2 \langle \lvert \hat{d} \rvert^2 \rangle_{\mathrm{FS}}}{T_c} \right) - \gamma_n T_c = \frac{\Delta_0^2 \langle \lvert \hat{d} \rvert^2 \rangle_{\mathrm{FS}}}{T_c} \geqslant 0,
    \label{eq:HeatJumpSingle}
\end{aligned}
\end{equation}
where $\gamma_n \equiv 4k_B^2 \zeta(2) \langle 1 \rangle_{\mathrm{FS}}$, with $\zeta$ being the Riemann zeta function, is the Sommerfeld coefficient. This shows that the anomalous behaviour of a negative specific heat jump can not occur for a single-component non-unitary order parameter, but only when the onset of non-unitary occurs as a subleading transition with $T_{c2} < T_{c1}$, consistent with the result of Eq.~\eqref{eq:Jump}.

%
\section{Validation of the topological phases}
\label{app:topo_validation}
%
To identify the topology of the nodal phases, we analyze slices with fixed momentum $k_x$ on which the Hamiltonian is gapped, except at the nodal point. The topology of the slices is identified by deforming the Hamiltonian into canonical form. The canonical form is a massive Dirac theory, for which the topology and its corresponding anomalous boundary excitations can be identified from an analysis of its symmetry-breaking mass terms \cite{GeierPhysRevB2018May,TrifunovicPhysRevX2019Jan}. The topology of the gapped phases is identified similarly by considering the Hamiltonian defined on the whole BZ. 

\subsection{Nodal phases}
{\em ${\rm B}_{3 {\rm u}}$ superconducting order.---}
The Bogoliubov-de Gennes Hamiltonian [Eq.~\eqref{eq:Hamk}] for a spherical Fermi surface with ${\rm B}_{3 {\rm u}}$ superconducting order parameter of the form $\vec{d}(\bo{k}) = (c_1 k_x k_y k_z, c_2 k_z, c_3 k_y)^{\mathrm{T}}$ can be written using Pauli matrices $\sigma$ in spin and $\tau$ in particle-hole space as
\begin{equation}
    H(\bo{k}) = \xi(\bo{k}) \sigma_0 \tau_3 - c_1 k_x k_y k_z \sigma_3 \tau_1 - c_2 k_z \sigma_0 \tau_2 + c_3 k_y \sigma_1 \tau_1. \, 
\end{equation}
The Hamiltonian has nodes at $\bo{k} = (\pm k_F, 0, 0)^{\rm T}$. Expanding to lowest order around the nodes, the Hamiltonian has the form of a massive Dirac theory
\begin{equation}
    H(\bo{k}) = m(k_x) \sigma_0 \tau_3 - c_2 k_z \sigma_0 \tau_2 + c_3 k_y \sigma_1 \tau_1, \, 
    \label{eq:app_H_nodal_B3u}
\end{equation}
with mass $m(k_x) = \xi((k_x, 0, 0)^{\rm T})$. 
With $\mathrm{B_{3u}}$ symmetric superconducting order, the Hamiltonian satisfies the symmetries
\begin{align}
    H(k_x, k_y, k_z) & = - U_{\cal P} H^*(-k_x, -k_y, -k_z) U_{\cal P}^\dagger \\ \nonumber
    & = U_{\cal T} H^*(-k_x, -k_y, -k_z) U_{\cal T}^\dagger \\ \nonumber
    & = U_{{\cal I}} H(-k_x, -k_y, -k_z) U_{{\cal I}}^\dagger \\ \nonumber
    & = U_{{\cal R}_x} H(k_x, -k_y, -k_z) U_{{\cal R}_x}^\dagger \\ \nonumber
    & = U_{{\cal R}_y} H(-k_x, k_y, -k_z) U_{{\cal R}_y}^\dagger \\ \nonumber
    & = U_{{\cal R}_z} H(-k_x, -k_y, k_z) U_{{\cal R}_z}^\dagger, 
\end{align}
with the representations $$U_{\cal P} = \sigma_0 \tau_1,$$ of particle-hole antisymmetry, $$U_{\cal T} = i \sigma_2 \tau_0,$$ of time-reversal symmetry, $$U_{\cal I} = \sigma_0 \tau_3,$$ of inversion symmetry, 
\begin{align*}
    U({\cal R}_{x}) & =i\sigma_{1}\tau_{3}, \\ 
    U({\cal R}_{y}) & =i\sigma_{2}\tau_{3}, \\ 
    U({\cal R}_{z}) & =i\sigma_{3}\tau_{0},
\end{align*}
of the rotation symmetries, as well as combinations thereof, in particular the chiral antisymmetry ${\cal C} = {\cal TP}$ with representation $$U_{\cal C} = U_{\cal T} U_{\cal P}^* = i \sigma_2 \tau_1,$$ and mirror symmetries ${\cal M}_j = {\cal IR}_j$, $j = x,y,z$ with representations 
\begin{align*}
    U({\cal M}_{x}) & =i\sigma_{1}\tau_{0}, \\
    U({\cal M}_{y}) & =i\sigma_{2}\tau_{0}, \\
    U({\cal M}_{z}) & =i\sigma_{3}\tau_{3}.
\end{align*}
The representations follow from the normal-state representations of spinful fermions and the symmetry of the superconducting order parameter \cite{GeierPhysRevB2020Jun}. 

The topological properties of a massive Dirac theory [such as Eq.~\eqref{eq:app_H_nodal_B3u}] can be obtained from an analysis of its mass terms~\cite{ChiuEA16, GeierPhysRevB2018May,TrifunovicPhysRevX2019Jan} and their behavior under the symmetries of the system. The mass terms are constant terms that anticommute with the linear-in-momentum terms of the Hamiltonian and anticommute mutually. Additional symmetry-allowed mass terms beyond the term proportional to $\xi(\bo{k})$ would allow to adiabatically deform the Hamiltonian to the topologically trivial form $H_{\rm triv.} = m \sigma_0 \tau_3$. 

To determine the properties of the nodal topological phase with nodes along the $k_x$-axis, we analyse the topological properties of the Hamiltonian defined on slices with fixed $k_x$ in the BZ.  Taking $k_x$ as a parameter in Hamiltonian Eq.~\eqref{eq:app_H_nodal_B3u} yields a massive Dirac theory whose additional mass terms are $$M_1 = m \sigma_{2}\tau_{1},$$ and $$M_2 = m \sigma_{3}\tau_{1}.$$
At $k_x = 0$, these two mass terms are prohibited by particle-hole antisymmetry ($M_2$), time-reversal symmetry ($M_1$ and $M_2$), chiral antisymmetry ${\cal C} = {\cal TP}$ ($M_1$), inversion symmetry ($M_1$ and $M_2$), and mirror symmetries ${\cal M}_x$ ($M_1$ and $M_2$) and ${\cal M}_y$, ${\cal M}_z$ ($M_2$). Slices with finite $0 < \lvert k_x \rvert < \pi$ satisfy a reduced set of symmetries. In this case, the mass terms are prohibited by chiral antisymmetry ${\cal C} = {\cal TP}$ ($M_1$), inversion-particle hole antisymmetry ($M_1$), and mirror symmetries  ${\cal M}_y$, ${\cal M}_z$ ($M_2$). The result that both additional mass terms $M_1$ and $M_2$ are prohibited by symmetries for slices $0 < \lvert k_x \rvert < \pi$ indicates that the massive Dirac theory Eq.~\eqref{eq:app_H_nodal_B3u} describes a topological phase with a topological invariant that protects the nodal point. 

The behavior of the (symmetry-forbidden) mass terms under the symmetries of the system determines the properties of the anomalous surface states~\cite{GeierPhysRevB2018May, TrifunovicPhysRevX2019Jan}. For $0 < \lvert k_x \rvert < \pi$, the result that the mass term $M_2 = \sigma_{3}\tau_{1}$ is prohibited only by mirror ${\cal M}_y$ and ${\cal M}_z$ symmetries and inversion-particle-hole antisymmetry ${\cal IP}$ indicates that the slices for $0 < \lvert k_x \rvert < k_F$ are in a second-order topological phase with zero-energy hinge states at mirror symmetric hinges~\cite{GeierPhysRevB2018May,TrifunovicPhysRevX2019Jan}. The chiral antisymmetry ${\cal PT}$ pins the flat hinge band to zero energy. At $k_x = 0$, the slice is furthermore invariant under time-reversal symmetry $U({\cal T}) = i \sigma_2 \tau_0 K$ and inversion symmetry $U({\cal I}) = \sigma_0 \tau_3$. At this point, the mass term $M_2 = m \sigma_{3}\tau_{1}$ is forbidden additionally by time-reversal symmetry and mirror symmetry ${\cal M}_x$. This indicates that the topological phase at $k_x = 0$ is first order with helical Majorana edge states and can be characterized by the topological invariant of a 2D topological superconductor in class DIII or a mirror Chern number. 

{\em ${\rm B}_{3 {\rm u}} + i \varepsilon {\rm B}_{2 {\rm u}}$ superconducting order.---} Admixture of a weak ${\rm B}_{2 {\rm u}}$ symmetric order with relative phase breaks rotation ${\cal R}_x$, ${\cal R}_y$, mirror ${\cal M}_x$, ${\cal M}_y$, and time-reversal symmetry. In this case, at $k_x = 0$, the mass terms are forbidden by inversion symmetry ($M_1$ and $M_2$), mirror ${\cal M}_z$ symmetry ($M_2$) and particle-hole antisymmetry ($M_2$). This indicates a second-order topological phase with zero-energy Majorana modes at mirror ${\cal M}_z$ symmetric hinges. At $0 < \lvert k_x \rvert < \pi$, the mass term $M_1$ is forbidden by the combination of inversion and particle-hole antisymmetry and the mass term $M_2$ is forbidden by mirror ${\cal M}_z$ symmetry. This corresponds to an obstructed atomic limit which does not have gapless boundary states but protects a nodal manifold around $(\pm k_F, 0, 0)^{\rm T}$. These boundary signatures are characteristic for a second-order topological nodal superconductor of type (iii)~\cite{SimonPRB2022}.

The fourfold degenerate node with $\mathrm{B_{3u}}$ pairing is split into two Weyl nodes with charge $\pm 1$. Each Weyl node describes a linear crossing of two eigenvalues which together with the symmetry analysis suffices for their identification as Weyl nodes from exact diagonalization of the Hamiltonian. 

{\em ${\rm B}_{3 {\rm u}} + i \varepsilon {\rm A}_{\rm u}$ superconducting order.---} Including an infinitesimal order parameter with ${\rm A}_{\rm u}$ symmetry breaks rotation ${\cal R}_{y}$ and ${\cal R}_{z}$, mirror ${\cal M}_{y}$ and ${\cal M}_{z}$, and, in case of a relative phase of the two superconducting orders, time-reversal symmetry. At $k_x = 0$, both mirror symmetry ${\cal M}_x$ and inversion symmetry ${\cal I}$ prohibit the mass terms $M_1$ and $M_2$. The slice at $k_x = 0$ has a non-trivial mirror Chern number. At $0 < \lvert k_x \rvert < \pi$, only the mass term $M_1$ is forbidden by the combination of inversion and particle-hole antisymmetry. Since $M_2$ is allowed, the slices with $0 < \lvert k_x \rvert < \pi$ can be trivialized which implies that the nodes at $(\pm k_F, 0, 0)^{\rm T}$ may generically gap out or split into multiple Weyl nodes with cancelling total charge as in our example in Sec.~\ref{sec:SpecificHeat}. 

{\em ${\rm B}_{3 {\rm u}} + i {\rm A}_{\rm u}$ superconducting order with $\vec{d}(\bo{k})=(0,c_{1}k_{z}+ic_{2}k_{y},0)^{{\rm T}}$.---} For a mixture $\vec{d}(\bo{k})=(0,c_{1}k_{z}+ic_{2}k_{y},0)^{{\rm T}}$, the Bogoliubov-de Gennes Hamiltonian is of the form
\begin{equation}
    H(\bo{k})= \xi(\bo{k}) \sigma_{0}\tau_{3}+c_{1}k_{z}\sigma_{0}\tau_{2}-c_{2}k_{y}\sigma_{0}\tau_{1}.
\end{equation}
This Hamiltonian describes a Weyl superconductor with fourfold degenerate Weyl nodes at $\bo{k}=(k_{F},0,0)^{{\rm T}}$. On slices with $\xi(\boldsymbol{k}) < 0$ ($\xi(\boldsymbol{k}) > 0$) , the system has Chern number $\mathrm{Ch}=2$ ($\mathrm{Ch}=0$). This Hamiltonian has ${\rm SU}(2)$ spin-rotation symmetry enforcing a two-fold spin degeneracy of the eigenstates.

\subsection{Gapped phases}
{\em ${\rm A}_{\rm u}$ superconducting order.---}
A fully gapped ${\rm A}_{\rm u}$ superconducting order with $\vec{d}(\bo{k}) = (c_1 k_x, c_2 k_y, c_3 k_z)^{\mathrm{T}}$ for a single band spherical Fermi surface has a Bogoliubov-de Gennes Hamiltonian of the form
\begin{equation}
    H(\bo{k})=\xi(\bo{k})\sigma_{0}\tau_{3}-c_{1}k_{x}\sigma_{3}\tau_{1}-c_{2}k_{y}\sigma_{0}\tau_{2}+c_{3}k_{z}\sigma_{1}\tau_{1}.
\end{equation}
This Hamiltonian describes a strong topological superconductor in class DIII with surface Majorana Dirac cones. It has a single mass term $M=m\sigma_{2}\tau_{1}$ that is prohibited by time-reversal $U(\mathcal{T})=i\sigma_{2}\tau_{0}K$ as well as by mirror symmetry $U({\cal M}_{x})=i\sigma_{1}\tau_{0}$. 

{\em ${\rm A}_{\rm u} + i \varepsilon {\rm B}_{3 {\rm u}}$ superconducting order.---} Breaking time-reversal symmetry by including an infinitesimal order with ${\rm B}_{3 {\rm u}}$ symmetry in addition to an ${\rm A}_{\rm u}$ superconducting order turns the system into a second-order topological superconductor with chiral Majorana modes on hinges preserving the ${\cal M}_{x}$ mirror symmetry.

\bibliography{Refs}
\end{document}